\pdfoutput=1
\documentclass[a4paper,12pt]{article}
\usepackage{a4}
\usepackage{epsfig}
\usepackage{latexsym, amssymb} 
\usepackage{amsmath}
\usepackage{psfrag}
\usepackage{bm}
\usepackage{feynmf}
\usepackage{rotating}
\usepackage{hyperref}
\usepackage{appendix}
\usepackage[utf8]{inputenc}
\hypersetup{
    unicode=false,          
    pdftoolbar=true,        
    pdfmenubar=true,        
    pdffitwindow=false,     
    pdfstartview={FitH},    
    pdftitle={My title},    
    pdfauthor={Author},     
    pdfsubject={Subject},   
    pdfcreator={Creator},   
    pdfproducer={Producer}, 
    pdfkeywords={keyword1} {key2} {key3}, 
    pdfnewwindow=true,      
    colorlinks=true,        
    linkcolor=red,          
    citecolor=blue,         
    filecolor=magenta,      
    urlcolor=green,         
    linktocpage=true
}
\def\beq{\begin{equation}}
\def\eeq{\end{equation}}
\def\br{\begin{eqnarray}}
\def\er{\end{eqnarray}}
\def\benu{\begin{enumerate}}
\def\efnu{\end{enumerate}}

\def\l{\left}
\def\r{\right}

\usepackage{authblk}

\title{\bf One spectrum to cure them all: Signature from early Universe solves major anomalies and tensions in cosmology}
\author[1,2,3]{Dhiraj Kumar Hazra~\footnote{dhiraj@imsc.res.in}}
\author[1,2]{Akhil Antony~\footnote{akhilantony@imsc.res.in}}
\author[4,5]{Arman Shafieloo~\footnote{shafieloo@kasi.re.kr}}
\affil[1]{\small The Institute of Mathematical Sciences, HBNI, CIT Campus, Chennai 600113, India}
\affil[2]{\small Homi Bhabha National Institute, Training School Complex, Anushakti Nagar, Mumbai 400085, India.}
\affil[3]{\small INAF/OAS Bologna, Osservatorio di Astrofisica e Scienza dello Spazio, Area della ricerca CNR-INAF, via Gobetti 101, I-40129 Bologna, Italy}
\affil[4]{\small Korea Astronomy and Space Science Institute, Daejeon 34055, Korea}
\affil[5]{\small University of Science and Technology, Daejeon 34113, Korea}
\date{\today}

\begin{document}

\maketitle

\abstract{Acoustic peaks in the Cosmic Microwave Background (CMB) temperature spectrum as observed by the Planck satellite appear to be smoother than our expectation from the standard model lensing effect. This anomalous effect can be also mimicked by a spatially closed Universe with a very low value of Hubble constant that consequently aggravates the already existing discordance between cosmological observations. We reconstruct a signature from the early Universe, a particular form of oscillation in the primordial spectrum of quantum fluctuations with a characteristic frequency, that solves all these anomalies. Interestingly, we find this form of the primordial spectrum resolves or substantially subsides, various tensions in the standard model of cosmology in fitting different observations, namely Planck CMB, clustering and weak lensing shear measurements from several large scale structure surveys, local measurements of Hubble constant, and recently estimated age of the Universe from globular clusters. We support our findings phenomenologically, by proposing an analytical form of the primordial spectrum with similar features and demonstrate that it agrees remarkably well with various combinations of cosmological observations. We support further our findings theoretically, by introducing a single scalar field potential for inflation that can generate such a form of the primordial spectrum.}  
\newpage
\tableofcontents
\section{Summary}
The standard model of cosmology is largely supported by the Cosmic Microwave Background (CMB) data from the Planck observation~\cite{Planck:2018param}. However, underneath this largely consistent picture, certain anomalies and tensions emerge as crucial problems of modern cosmology. 

Excess smearing of the acoustic peaks in the CMB temperature power spectrum compared to the lensing effect predicted from the standard model creates an \textit{anomaly}. A similar temperature power spectrum is also obtained in a spatially curved Universe but with very small values of Hubble constant that drastically changes the expansion history and that in turn aggravates an already existing strong tension with local Hubble measurements~\cite{Riess:2019,H0LiCOW:2019}. Moreover, Planck data prefers a higher value for the combination of matter density and growth of structure parameters, known as $S_8$, which is not consistent with lensing measurements~\cite{DES:2017,KiDS1000:Asgari}.

As a solution to these anomalies, we propose a signature from the early Universe, specifically in the spectrum of primordial quantum fluctuation. Importantly, we demonstrate that a solution to the anomalies within the Planck CMB data automatically resolves or greatly reduces the tension between different datasets and restores the cosmic concordance. 

Being a representative of \textit{new physics}, our solution will also encourage new models of high energy physics. With upcoming observations, the primordial signature can be tested with better precision.

\section{Models, data, anomalies and tensions}~\label{sec:anomalies}

\paragraph{The Standard Model:} The {\tt Standard Model} discussed in this article refers to the spatially flat $\Lambda$ (\textit{Cosmological Constant} as dark energy) CDM (non-interacting or~\textit{cold} dark matter) Universe with power law form of primordial power spectrum. The power spectrum is defined with an amplitude ($A_s$) and tilt ($n_s$) at a pivot scale ($k_{pivot}$) as $\mathcal{P}_{power~law}(k)=A_s\l(k/k_{pivot}\r)^{n_s-1}$. This model is same as the \textit{baseline} model used in Planck CMB analysis.

In this article, we will mainly discuss the analysis with Planck CMB data. For extended analyses we utilize the small scale CMB from Atacama Cosmology Telescope (ACT)~\cite{ACT:2020} and lensing and clustering data from the Dark Energy Survey (DES)~\cite{DES:2017}. We also incorporate and discuss the analysis using Baryon Acoustic Oscillation (BAO) data ~\cite{BAO6DF:2011,BAODR12BOSS:2016} and supernovae data~\cite{Pantheon:2018}. The likelihood and the data that we use, are described in~\autoref{sec:methods} and listed in~\autoref{tab:data}. 

While the {\tt Standard Model} of cosmology is supported by the CMB and different large scale structure datasets, certain statistically significant inconsistencies in its fits to the data remain as the crucial problems of modern cosmology. These inconsistencies can be divided into two categories, \textit{anomalies and tensions}.

\paragraph{Anomalies and Tensions:} By anomalies here, we refer to internal inconsistencies within the context of the {\tt Standard Model} fitting the Planck CMB data. We discuss here two major anomalies found in the CMB temperature anisotropy data from Planck observation, namely the lensing amplitude and the curvature. By tensions we refer to inconsistencies in the posterior inference of the key cosmological parameters such as Hubble constant, amplitude of the growth of structure $\sigma_8$, and age of the Universe, obtained upon fitting the {\tt Standard Model} to different independent data including CMB, weak lensing, baryon acoustic oscillation (BAO) and supernovae data.

\paragraph{Lensing amplitude ($A_{lens}$) Anomaly:}
Lensing of the CMB anisotropies became evident with Planck and other small scale CMB measurements. While the lensing potential power spectrum measurement provides a direct estimation of lensing, the effect of lensing is also strongly manifested as smearing of acoustic peaks. In a {\tt Standard Model} the theoretical prediction of lensing can be calculated directly, given the background cosmology and the initial condition. Therefore apart from the standard 6 parameter model, lensing does not require any extra parameter. However, as a consistency check an \textit{ad hoc} parameter $A_{lens}$ is assumed as an amplitude parameter that scales the theoretical prediction for lensing power spectrum. For the {\tt Standard Model}, $A_{lens}=1$. When $A_{lens}$ is allowed to vary in an analysis with Planck temperature angular power spectrum data, it is found to be higher than 1 ($A_{lens}=1.243\pm0.096$ from P18TT and $A_{lens}=1.180\pm0.065$ from  P18TP), indicating excess lensing, breaking the consistency condition. Since this excess lensing, obtained as an \textit{unphysical} multiplicative factor, it is categorized as an \textit{anomaly}. It should be noted that the temperature-polarization cross spectrum prefers (not significantly though) a lower lensing amplitude. Recent small scale CMB data from the observation with Atacama Cosmology Telescope~\cite{ACT:2020} finds complete consistency, with $A_{lens}$ being consistent to unity. Therefore, at present, the lensing anomaly can also be viewed as a disagreement between two observations of the same nature.
\paragraph{Curvature ($\Omega_k$) Anomaly:}
In a curved Universe, the bending of photons imprints a lensing effect. The article~\cite{DiValentino:2019} demonstrates the positive correlation between the curvature and lensing amplitude and has argued that a closed Universe can solve the lensing anomaly problem. Since curvature changes the distance measure, it shifts the position of the acoustic peaks (curvature also has a marginal effect in lowering the power at very large multipoles.). Lowering the value of Hubble constant ($H_0=52.2\pm4.3$ from P18TT) helps in shifting back the peaks to the observed positions. This also results in a significant shift in matter density. These shifts make the estimated Hubble constant completely inconsistent with the local measurements ($H_0=74.03\pm1.42$~\cite{Riess:2019}). The increased matter density also becomes 3.5$\sigma$ inconsistent with the independent measurements from the cosmic shear and clustering observations. Therefore a closed Universe, that solves the lensing anomaly, aggravates the discordance between datasets. Apart from the observational point of view, high curvature density poses a challenge in the theory of inflation. Inflation predicts a flat Universe as the nearly exponential expansion reduces initial curvature by orders of magnitude. A closed Universe with a high curvature density $\Omega_k\sim{\cal O} (0.1)$ today would mean a large curvature density during the onset of inflation. That requires the initial condition on the scalar field that drives the inflation to be fine 
tuned. We term this as \textit{curvature anomaly} as it brings in problems in the theoretical construct in addition to strongly aggravating the disagreements between different datasets.


\paragraph{The Hubble tension:}
Hubble constant obtained from the {\tt Standard Model} fit to the Planck CMB data is estimated to be $H_0=66.88\pm0.92$ (from P18TT) and $67.27\pm0.60$  (from P18TP)~\cite{Planck:2018param}. Local measurement of Hubble constant from Cepheids and supernovae indicates a value significantly larger than these with $H_0=74.03\pm1.43$~\cite{Riess:2019}. Other measurements of the Hubble constant from time-delay cosmography of the lensed quasars~\cite{H0LiCOW:2019,Liao:2020zko} and calibration of the Tip of Red Giant Branch (TRGB)~\cite{CCHP:2020} also indicate Hubble constant to be at higher values, though these measurements have larger uncertainties and show less tension with the estimated value of the Hubble constant fitting the {\tt Standard Model} to Planck data.

\paragraph{Tension in amplitude of the growth of structure and $S_8$:}
Measurement of cosmic shear in lensing surveys constrains the matter density ($\Omega_m$) and the amplitude of linear matter power spectrum ($\sigma_8$). Since these two parameters have degenerate effects in the lensing surveys, usually $S_8=\sigma_8\sqrt{\Omega_m/0.3}$ is used as a parameter to compare the consistency with other observations. Clustering and shear analyses provides $S_8=0.772^{+0.018}_{-0.017}$~\cite{DES:2021} and  $S_8=0.759^{+0.024}_{-0.021}$~\cite{KiDS1000:Asgari}. Planck CMB estimates $S_8=0.840\pm0.024$ (P18TT) and $S_8=0.834\pm0.016$ (P18TP). The disagreement is evident and it can also be appreciated that the tension with the CMB is driven mainly by the temperature data.\\

 Anomalies in an experiment and observation and tensions between datasets are not uncommon. In certain cases, they are found to be originating from systematic effects. In other cases, these can turn out to be the indicators of \textit{new physics}. Usually, the solution to the anomalies and tensions are proposed \textit{separately} with different extensions to the {\tt Standard Model}. Mimicking lensing effect has been discussed by adding oscillations in the primordial spectrum~\cite{HazraMRLPlanck:2014,Planck:2018inf,Domenech:2019cyh,Domenech:2020qay}. These solutions do not address the tensions between the datasets. Early dark energy and late time transition in dark energy have been proposed as possible solutions to alleviate the tensions but none of these alternative models have been able to satisfy all different combinations of the data successfully~\cite{Poulin:2018,Agrawal:2019,Hill:2020,HillACT:2021,Poulin:2021bjr,Li:2019yem,Li:2020ybr,DiValentino:2020zio,DiValentino:2020vvd}. In fact larger degrees of freedom in most these models result in inflated confidence balls (without shifting much the mean values of the key parameters) that helps in overlapping of the contours. In most such approaches one could only reduce the significance of the tensions by inflated confidence limits rather than solving the problem. One should note that these solutions do not address the anomalies either. A possible solution with primordial spectrum in the context of Hubble tension has been proposed in~\cite{HazraHST:2018,Keeley:2020} by two of the authors of this article. However, finding a theoretical solution was not possible given the complicated shape of the spectrum obtained.\\

In this article we explore a common solution to the anomalies and tensions (in developing our methodology we take hints from~\cite{HazraMRLPlanck:2014} and~\cite{HazraHST:2018}). A common solution that solves both inconsistencies naturally enhances the chances of the solution being a possible candidate for new physics.

\section{Approach}~\label{sec:Approach}

We take a top down approach to solve the anomalies. We reconstruct the shape of the primordial power spectrum of quantum fluctuations directly from the data using deconvolution techniques. Upon comparing the reconstructed spectrum with the data, we show that the anomalies can be removed and a flat Universe remains completely consistent with the data without any excess lensing. 

The reconstructed spectrum is then parametrized by an analytical function. Comparing with different datasets we obtain the significance of the proposed spectrum. Here we use Markov Chain Monte Carlo and nested sampling techniques to compare with observational data from different sources. 

We finally propose a potential of the scalar field responsible for inflation, assuming inflation being the source of the primordial fluctuation. This potential provides a theoretical support for the proposed spectrum.

\section{Results}~\label{sec:Results}
Our results are mainly divided into two parts, reconstruction and analytical form of a new spectrum.
We point out main results of our analyses below, before the detailed discussions.
\paragraph{Summary of results:}

\begin{itemize}
    \item A simple shape of primordial power spectrum, obtained through deconvolution (hereafter referred as {\tt Reconstruction}) \textit{solves the lensing amplitude anomaly}. 
    \item The {\tt Reconstruction} also \textit{solves the closed Universe anomaly and brings back cosmic concordance}.
    \item Importantly, we find that a solution to the anomalies within the Planck CMB data automatically resolves or greatly reduces the tension between different datasets.
    
    \item Our analytical power spectrum model, {\tt New Spectrum}, that is designed to match the {\tt Reconstruction}, prefers lower matter density and lower $\sigma_8$ and higher $H_0$. When the {\tt New Spectrum} and {\tt Restricted Spectrum} (a simpler version of the former) are compared with Planck temperature data with priors on Hubble constant, we get moderate to very strong evidence for the models compared to the {\tt Standard Model}.
    
    \item The proposed form of the spectrum stays consistent with small scale CMB measurements from Atacama Cosmology Telescope observation~\cite{ACT:2020}, large scale structure measurements from Dark Energy Survey~\cite{DES:2017} and recently estimated age of the Universe from globular clusters~\cite{GlobularClusters:2020,GlobularClusters:2021}.

\end{itemize}

\subsection{Reconstructed spectrum and the concordance}~\label{sec:reconstruction}

\begin{figure*}[!htb]
\centering
\includegraphics[width=\columnwidth]{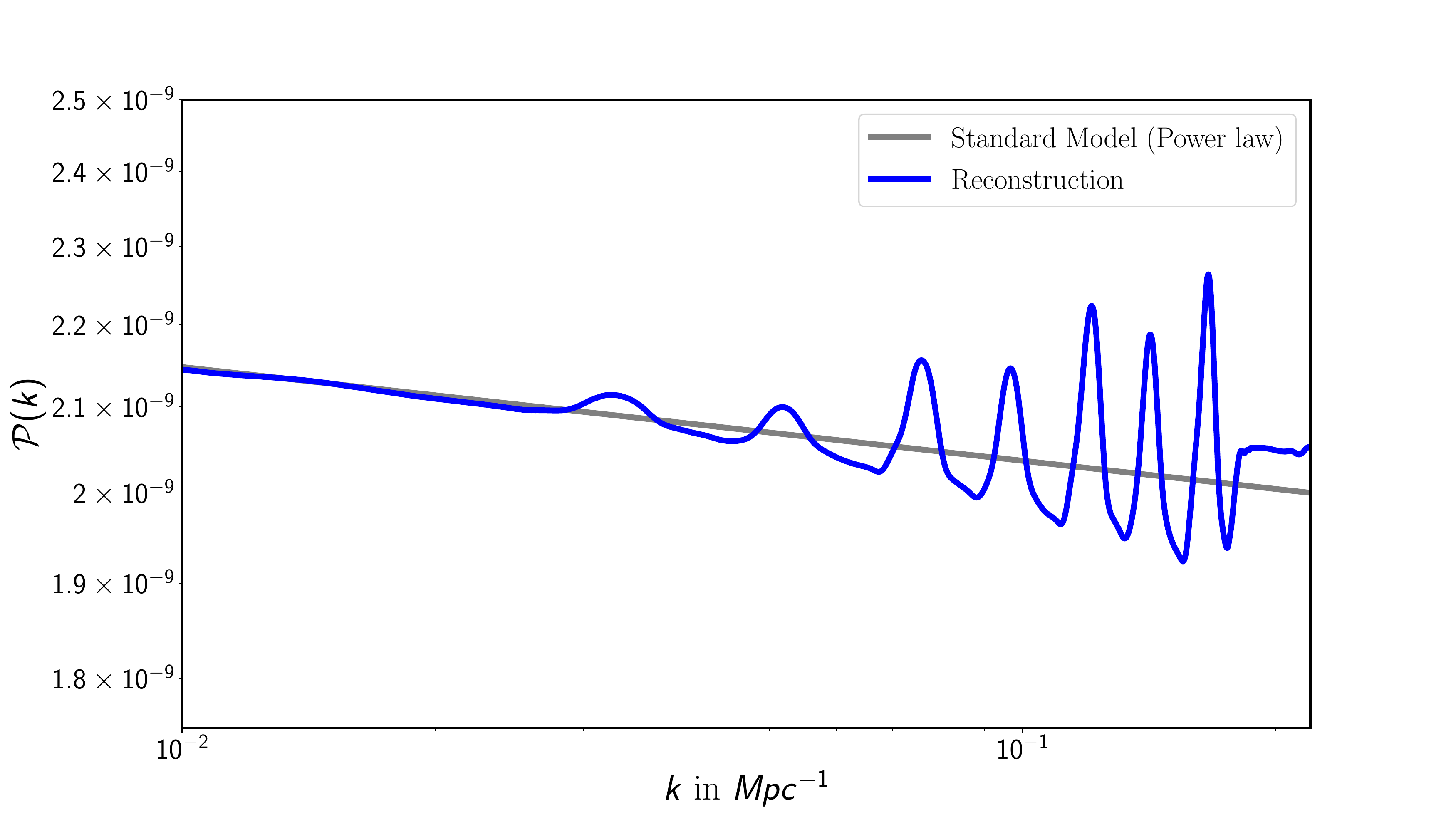}
\caption{\footnotesize\label{fig:Recon} {\bf Reconstructed spectrum that brings back cosmological concordance.} A smoothed version of the reconstructed spectrum is plotted with the power law form of primordial spectrum representing the {\tt Standard Model}.} 
\end{figure*}

We begin with presenting the reconstructed primordial power spectrum that mimics the best fit temperature power spectrum to the Planck temperature data in the {\tt Standard Model} with $A_{lens}$ extensions. A smoothed version of the reconstructed spectrum is plotted in~\autoref{fig:Recon} along with the power law form of the primordial power spectrum. This spectrum removes the lensing anomaly as it prefers $A_{lens}=1$ in fitting the Planck temperature data. At the same time, this spectrum prefers a flat Universe ($\Omega_k=0$).

In the next few plots, we will compare the results of our analysis using the reconstructed spectrum against the {\tt Standard Model} results with their extension. In this subsection we compare the reconstructed spectrum with the Planck temperature data only as the tension with other data and anomalies are mainly related to Planck temperature data with low-$\ell$ polarization data from High Frequency Instrument (P18TT).

\begin{figure*}[!htb]
\centering
\includegraphics[width=0.47\columnwidth]{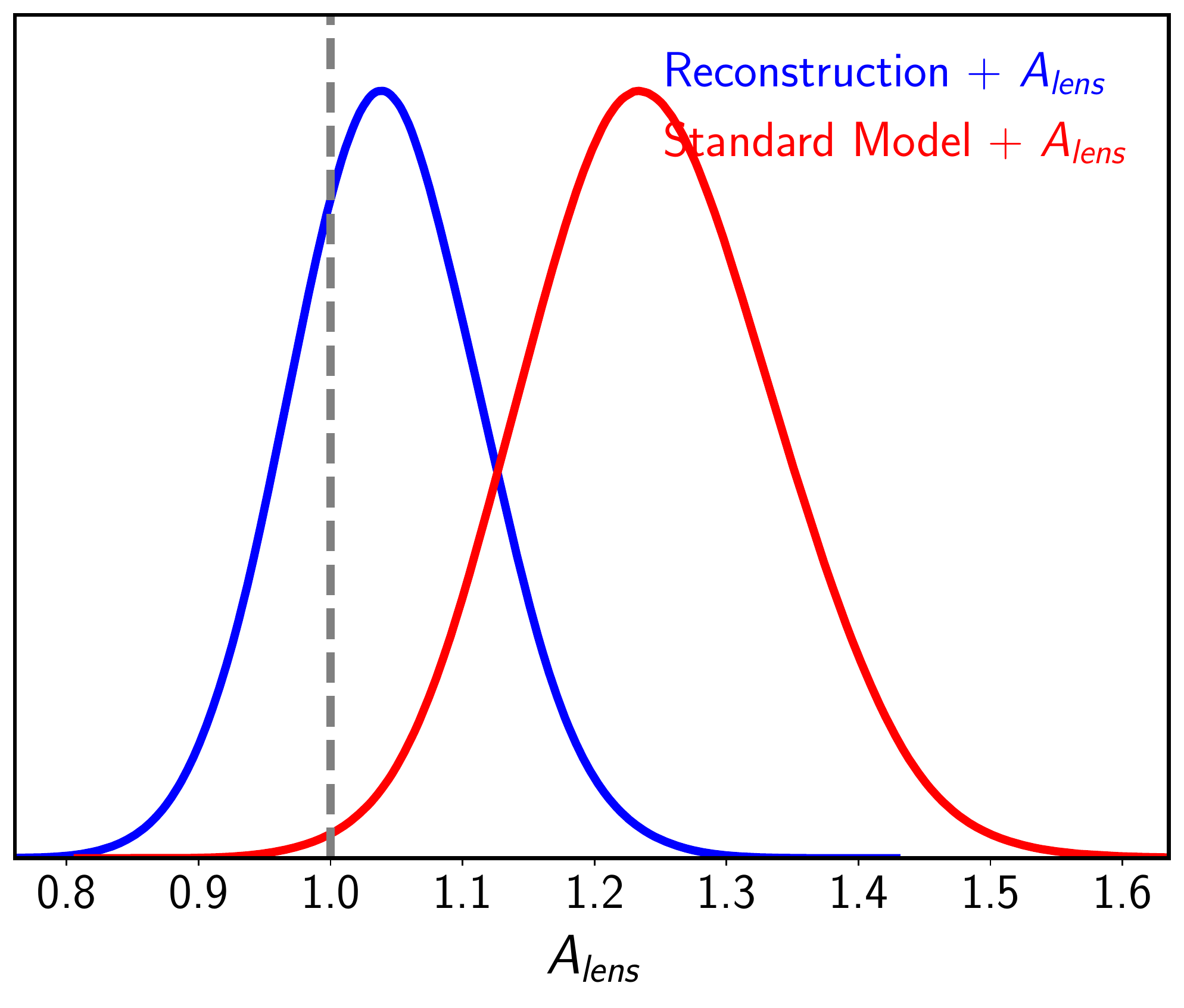}
\includegraphics[width=0.49\columnwidth]{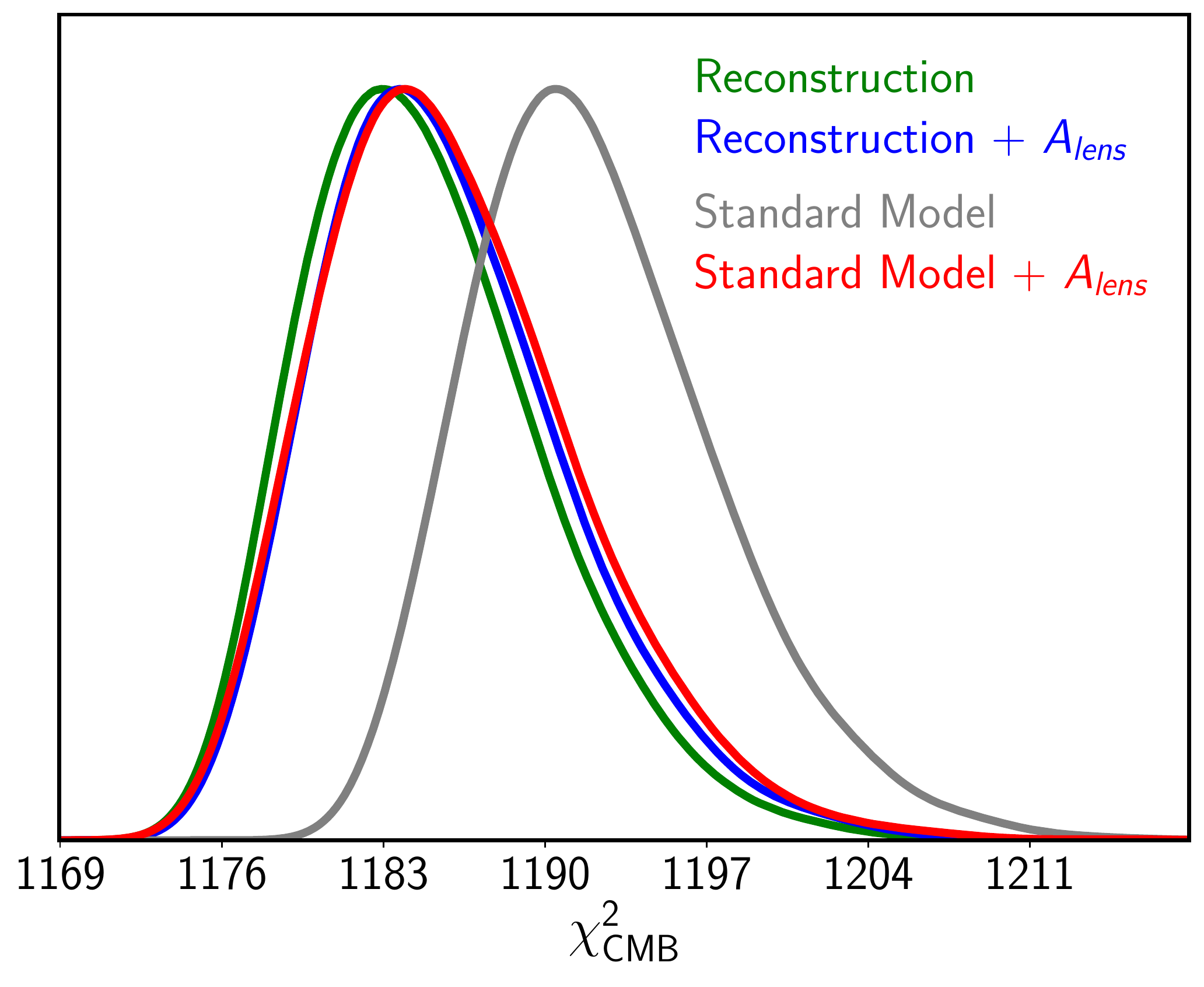}

\caption{\footnotesize\label{fig:ReconAlens} {\bf Solving the lensing anomaly.} [Left] Marginalized posteriors of lensing amplitude ($A_{lens}$) obtained in the {\tt Standard Model} and when reconstructed spectrum is used. [Right] Marginalized posteriors of $\chi^2$ from CMB in these two analyses compared with the {\tt Standard Model} where lensing amplitude is fixed to 1. Both analyses with {\tt Reconstruction} and {\tt Reconstruction +} $A_{lens}$ provide significant improvements in fit to the data compared to the {\tt Standard Model}. The posterior distributions of $\chi^2$ are also nearly indistinguishable from the distribution obtained in {\tt Standard Model +} $A_{lens}$ analysis.} 
\end{figure*}

In~\autoref{fig:ReconAlens} we plot the marginalized posteriors for the parameter $A_{lens}$ on the left. The {\tt Standard Model} where we use the power law form of the primordial spectrum prefers a larger value of $A_{lens}=1.243\pm 0.096$ ($1\sigma$ confidence interval). When reconstructed power spectrum is used, we find $A_{lens} = 1.042\pm 0.074$. Note that the mean value of  $A_{lens}$ reaches significantly closer to unity that solves the near-3$\sigma$ lensing anomaly existing in the {\tt Standard Model}. In the same plot, to the right we compare the distribution of $\chi^2$ in these three models. Note that {\tt Standard Model +} $A_{lens}$ provides significant improvement in fit to the data compared to the {\tt Standard Model}. The distributions obtained from {\tt Reconstruction} and {\tt Reconstruction +} $A_{lens}$ analyses are almost identical to the {\tt Standard Model +} $A_{lens}$. Both $A_{lens}$ and $\chi^2$ distributions indicate that the analysis with reconstructed spectrum provides very similar fit to the Planck data, remaining completely consistent with $A_{lens}=1$. We provide corresponding triangle plots in~\autoref{fig:ReconAlensTri} in the supplementary materials for further comparison of parameter constraints.

\begin{figure*}[!htb]
\centering
\includegraphics[width=0.50\columnwidth]{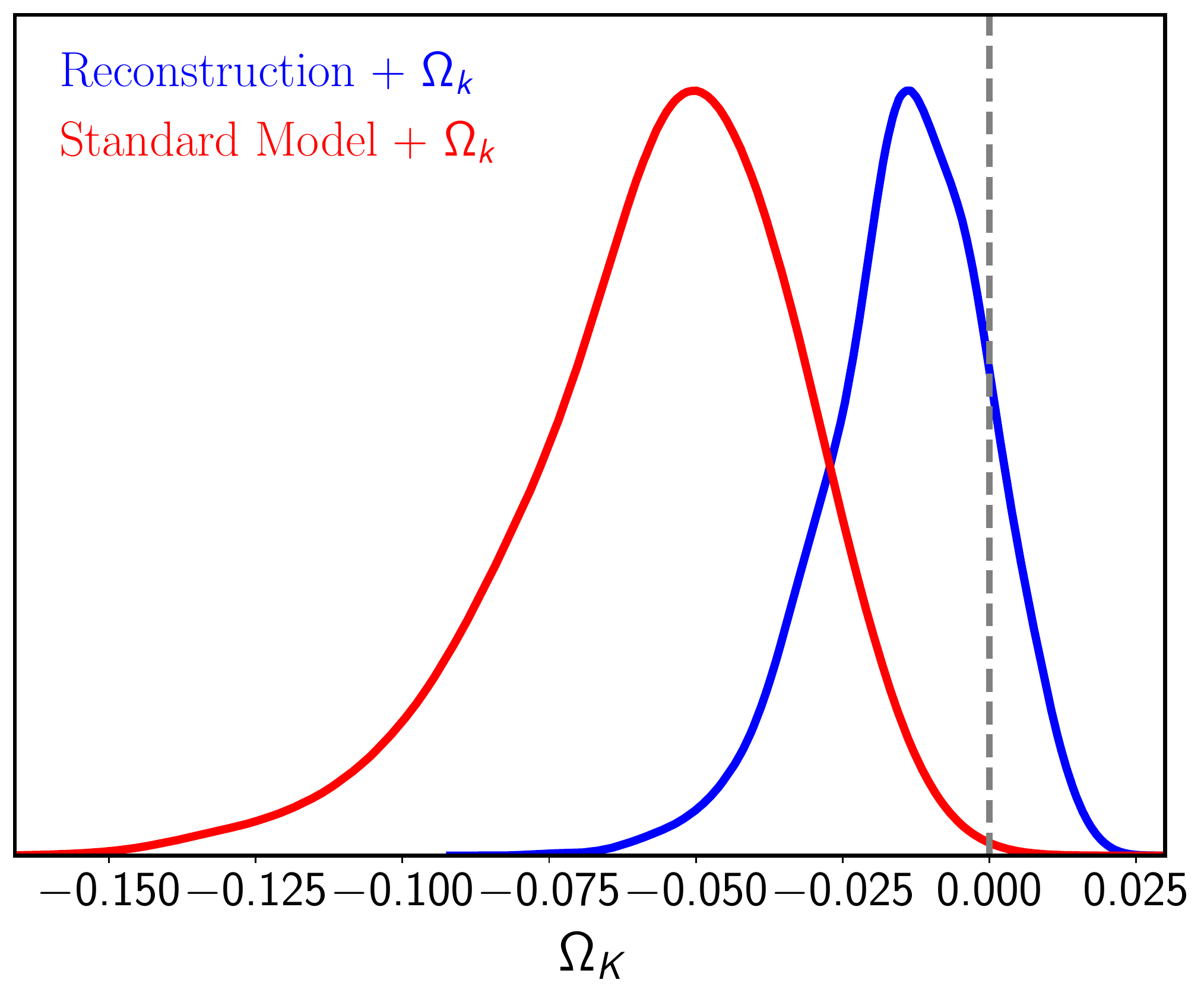}
\includegraphics[width=0.492\columnwidth]{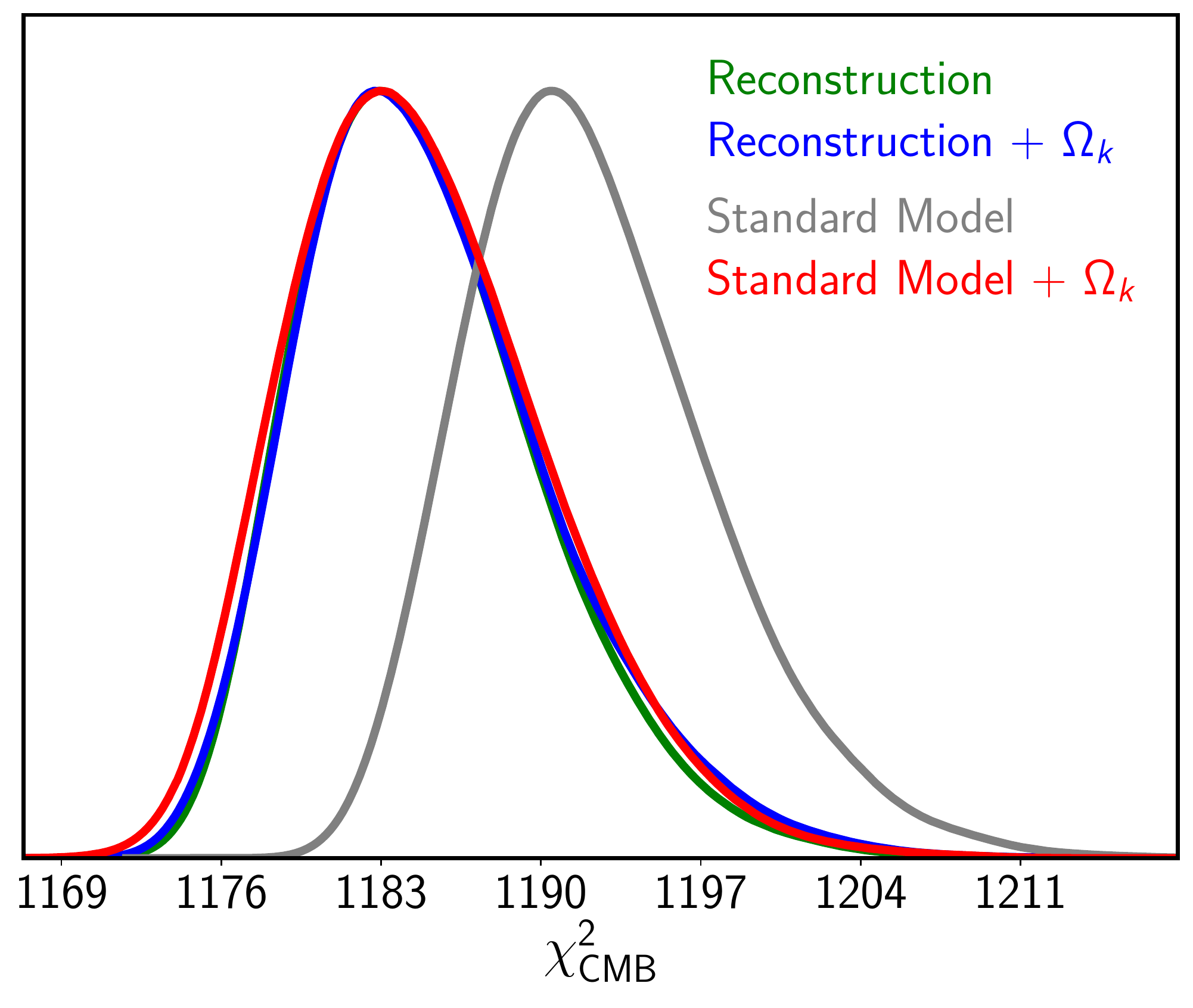}
\caption{\footnotesize\label{fig:ReconOmegak} {\bf Bringing the flat Universe back.} [Left] Marginalized posteriors of curvature density parameter $\Omega_K$ in the {\tt Standard Model} and when reconstructed spectrum is used. [Right] Marginalized posteriors of $\chi^2$ from CMB in these two analyses compared with the {\tt Standard Model} with flat Universe. Note that both analyses, {\tt Reconstruction} and {\tt Reconstruction +} $\Omega_K$ provide significant improvements in fit to the data compared to the {\tt Standard Model} and the distributions are nearly identical to the distribution obtained in {\tt Standard Model +} $\Omega_K$.} 
\end{figure*}

In order to demonstrate that the reconstructed power spectrum also mimics the effect of curvature of the Universe, we compare the analysis of {\tt Reconstruction +} $\Omega_K$ with {\tt Standard Model +} $\Omega_K$ in~\autoref{fig:ReconOmegak}. The posteriors are skewed and the bounds upto $3\sigma$ in {\tt Standard Model +} $\Omega_K$ is found to be $\Omega_K = -0.056^{+0.028+0.044+0.050}_{-0.018-0.050-0.079}$ that indicates little over 3$\sigma$ preference for a closed Universe from Planck temperature anisotropy data. {\tt Reconstruction +} $\Omega_K$ analysis finds $\Omega_K = -0.014^{+0.016}_{-0.011}$ that brings the flat Universe ($\Omega_K=0$) back within 68\% confidence interval. The posterior plots of $\Omega_K$ are provided on the left while the distributions of $\chi^2$ are plotted to the right. Similar to~\autoref{fig:ReconAlens}, we find here that the $\chi^2$ distribution is identical in {\tt Reconstruction}, {\tt Reconstruction +} $\Omega_K$ and {\tt Standard Model +} $\Omega_K$. Therefore when reconstructed spectrum is used as the primordial spectrum, a flat Universe provides a similar improvement in fit as the {\tt Standard Model} in a closed Universe. Corresponding triangle plots for parameter constraints are available in~\autoref{fig:ReconOmegaKTri} in the supplementary materials.

\begin{figure*}[!htb]
\centering

\includegraphics[width=0.6\columnwidth]{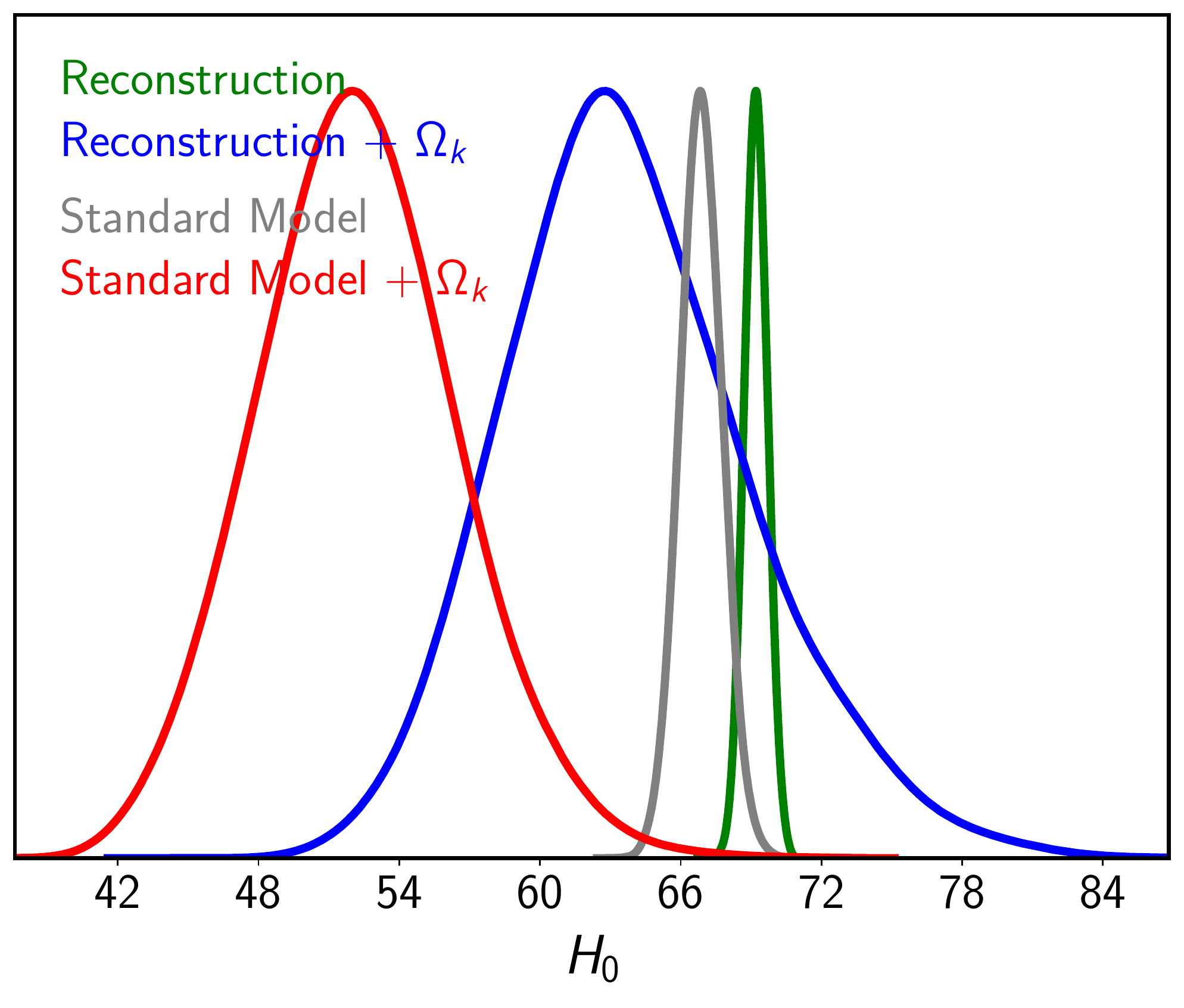}

\caption{\footnotesize\label{fig:ReconHubble} {\bf Determination of the Hubble parameter.} Marginalized posteriors of $H_0$ from extended model compared with the {\tt Standard Model}. A closed Universe preferred by Planck temperature data prefers a significantly lower value of the Hubble constant while Reconstructed power spectrum fits the Planck data with significantly larger values of $H_0$ solving the discordance.} 
\end{figure*}

\autoref{fig:ReconAlens} and~\autoref{fig:ReconOmegak} demonstrates a strong degeneracy between primordial physics, curvature and lensing. Note that here the most probable physics can only be identified by the theoretical support and consistency with other datasets. As discussed earlier, lensing amplitude is multiplied as an ad-hoc parameter to the theoretically determined lensing power spectrum. Therefore $A_{lens}$ is termed as a `consistency parameter'. On the other hand while curvature is an intrinsic property of the space-time, Planck~\cite{Planck:2018param} and other~\cite{DiValentino:2019} analyses point out the significantly lower Hubble constant preferred by a closed Universe. In~\autoref{fig:ReconHubble} we compare the posteriors on the Hubble constant obtained in different analyses. {\tt Standard Model +} $\Omega_K$ prefers a significantly lower value of Hubble constant, though with larger uncertainly due to strong degeneracy in the measurement of distances. $H_0 = 52.2\pm 4.3$ found in {\tt Standard Model +} $\Omega_K$ analysis while {\tt Reconstruction +} $\Omega_K$ prefers higher value with $H_0 =63.8^{+4.5}_{-5.8}$. 

\begin{figure*}[!htb]
\centering
\includegraphics[width=\columnwidth]{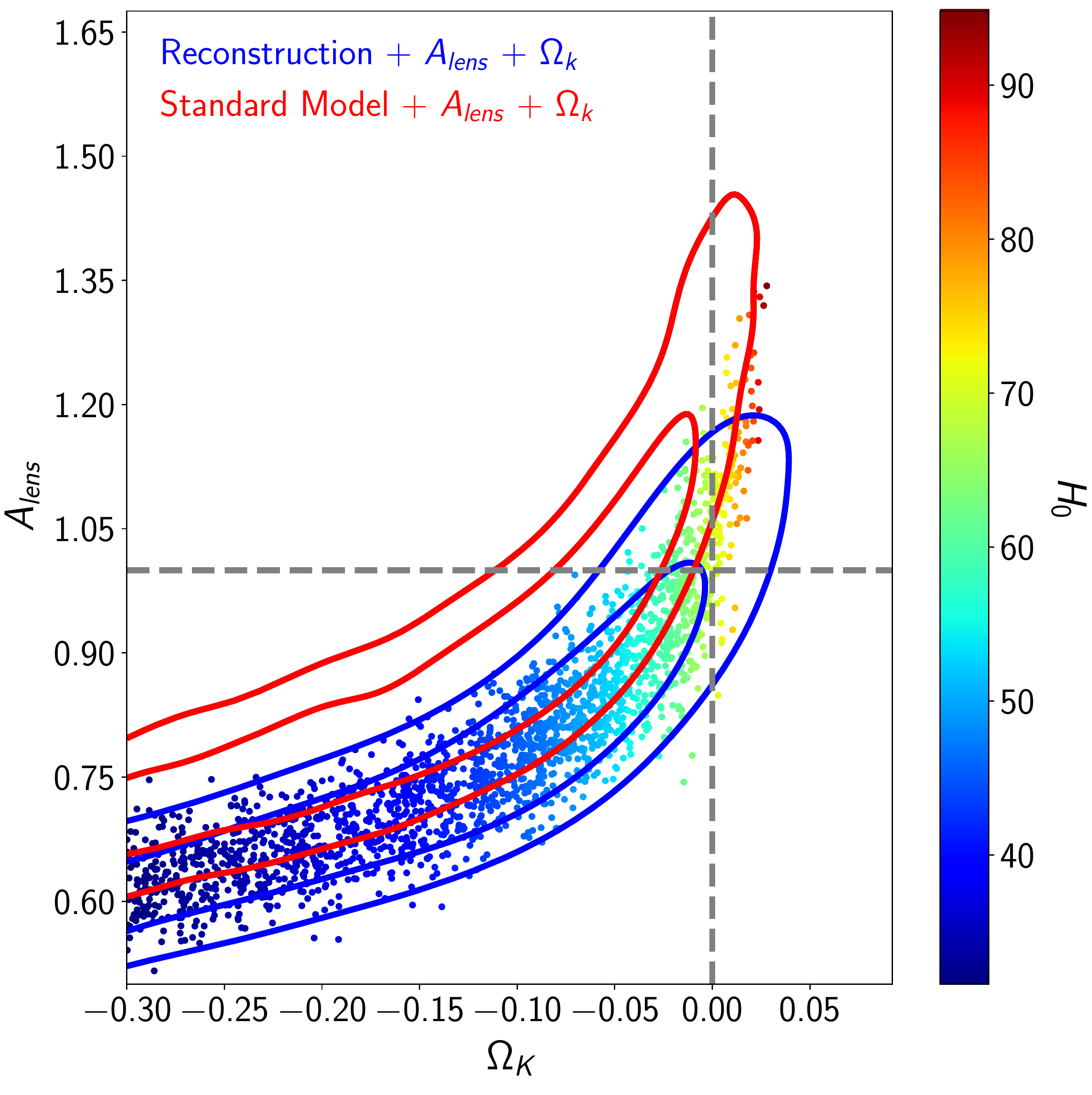}
\caption{\footnotesize\label{fig:ReconOmegakAlens} {\bf Bringing back the concordance.} 2D contour plot for lensing amplitude and curvature density parameters. Samples colored with Hubble constant are plotted for the analysis with reconstructed spectrum.}
\end{figure*}

The degeneracy between lensing amplitude and curvature has been discussed in detail in~\cite{DiValentino:2019}. A closed Universe solves lensing anomaly but increases the tension between other datasets on the determination of cosmological parameters. The $A_{lens}-\Omega_K$ degeneracy is plotted in~\autoref{fig:ReconOmegakAlens}. Note that here $\Omega_K$ becomes unbounded from below with $\Omega_K = -0.125^{+0.12+0.013}_{-0.068 <0.3}$ (uncertainties quoted upto $2\sigma$). Note that with the increased degeneracy, the preference for the closed Universe remains at 2$\sigma$. Reconstructed power spectrum, however, resolves both the anomalies. In blue, we have plotted 1-2$\sigma$ confidence bands on the same parameter pair in {\tt Reconstruction +} $A_{lens}$ + $\Omega_K$ analysis. Importantly, here the contours simply shift vertically down compared to the contours obtained from {\tt Standard Model +} $A_{lens}$ + $\Omega_K$, thus making the flat Universe consistent with the Planck temperature data with lensing amplitude of unity. We also plot the samples of $H_0$ in colormap. Samples indicate that in {\tt Reconstruction +} $A_{lens}$ $\Omega_K$ analysis, around $A_{lens}=1$ and $\Omega_K=0$ the model prefers higher values of Hubble constant ($H_0\sim70$) compared to the {\tt Standard Model}

\begin{figure*}[!htb]
\centering
\includegraphics[width=0.47\columnwidth]{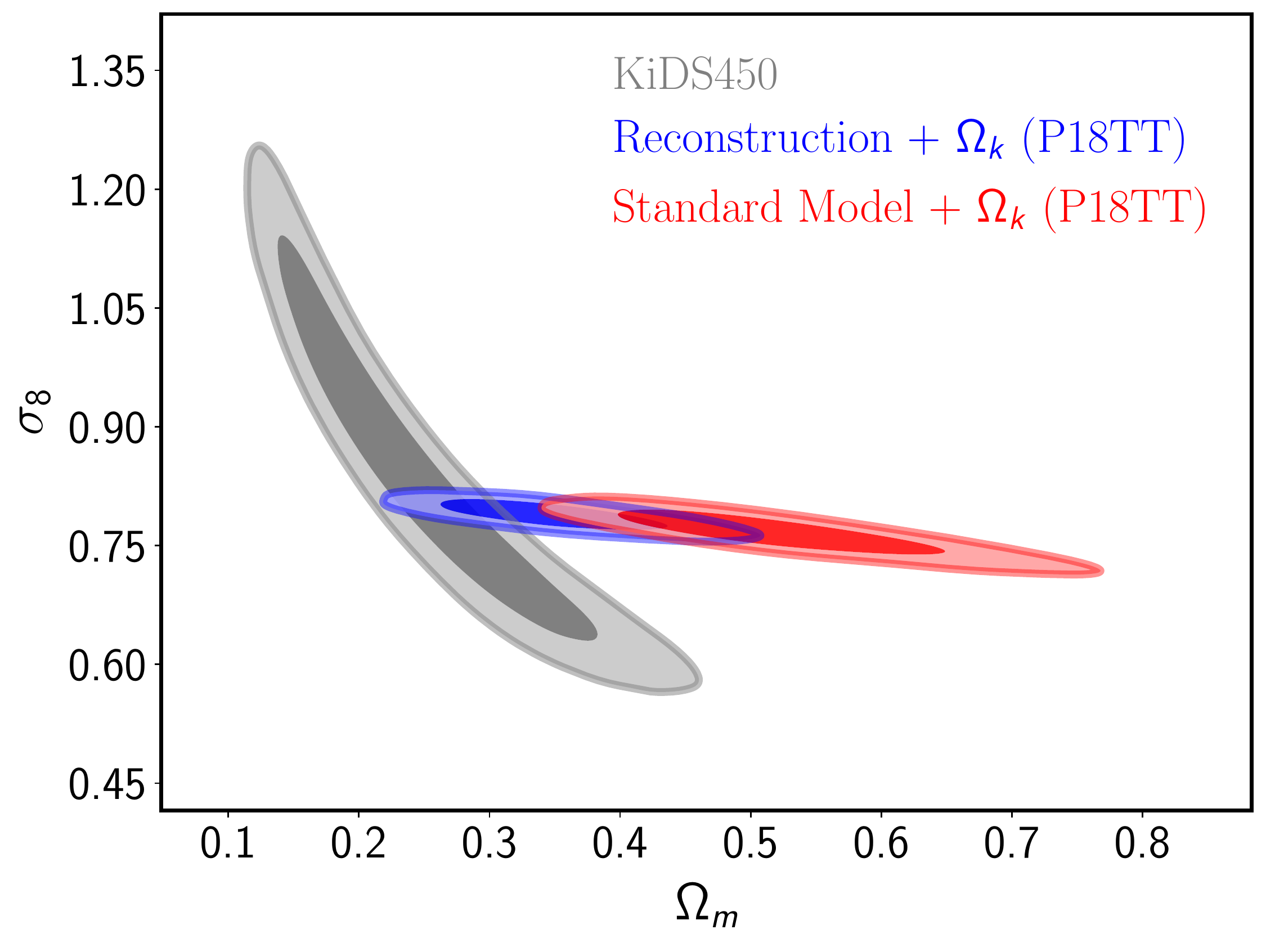}
\includegraphics[width=0.42\columnwidth]{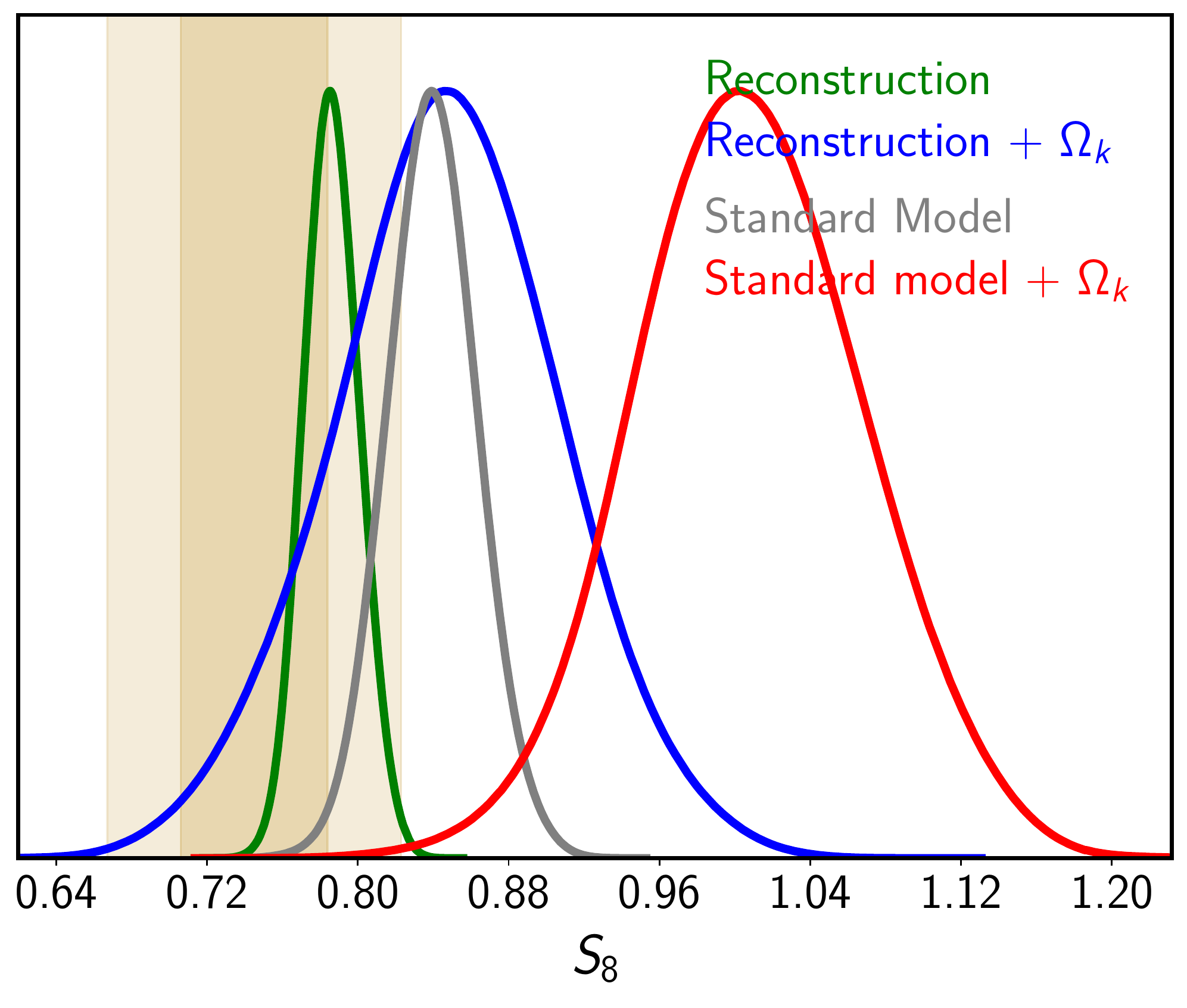}
\caption{\footnotesize\label{fig:ReconLSS1} {\bf Comparison with Large Scale Structure $\Omega_m-\sigma_8$.} Discordance between CMB and lensing surveys in the estimation of $\Omega_m-\sigma_8$. Note that a discordance over 3$\sigma$ in {\tt Standard Model +} $\Omega_K$ can be completely ameliorated when reconstructed spectrum is used as the primordial spectrum instead of power law form. The vertical band denotes the constraint from KiDS450.} 
\end{figure*}

Apart from a very low Hubble constant, a closed Universe also aggravates the discordance between CMB and lensing surveys in the estimation of $\Omega_m-\sigma_8$ parameter. In~\autoref{fig:ReconLSS1}, $\Omega_m-\sigma_8$ correlation is plotted in the left. The strong disagreement with KiDS450~\cite{KiDS450:Hidelbrandt} is evident in the {\tt Standard Model +} $\Omega_K$ analysis with Planck temperature data (see also~\cite{DiValentino:2019}). Interestingly, the {\tt Reconstruction +} $\Omega_K$ model laterally shifts the $\Omega_m-\sigma_8$ contour in the direction of lower matter density that brings back the concordance, providing substantial overlap. Marginalized posteriors of $S_8=\sigma_8\sqrt{\Omega_m/3}$ parameter are plotted in the right. Yellow band indicates the 1-2$\sigma$ bounds from KiDS450 ($S_8=0.745\pm0.039$) analysis~\cite{KiDS450:Hidelbrandt} (KiDS1000 analysis reports and updated value of $S_8=0.759^{+0.024}_{-0.021}$~\cite{KiDS1000:Asgari}). Note that while $S_8$ posteriors from {\tt Standard Model} is at more than $2\sigma$, when curvature is allowed to vary, the tension becomes more than 6$\sigma$. On the other hand reconstructed spectrum removes the tension.

\begin{figure*}[!htb]
\centering
\includegraphics[width=0.6\columnwidth]{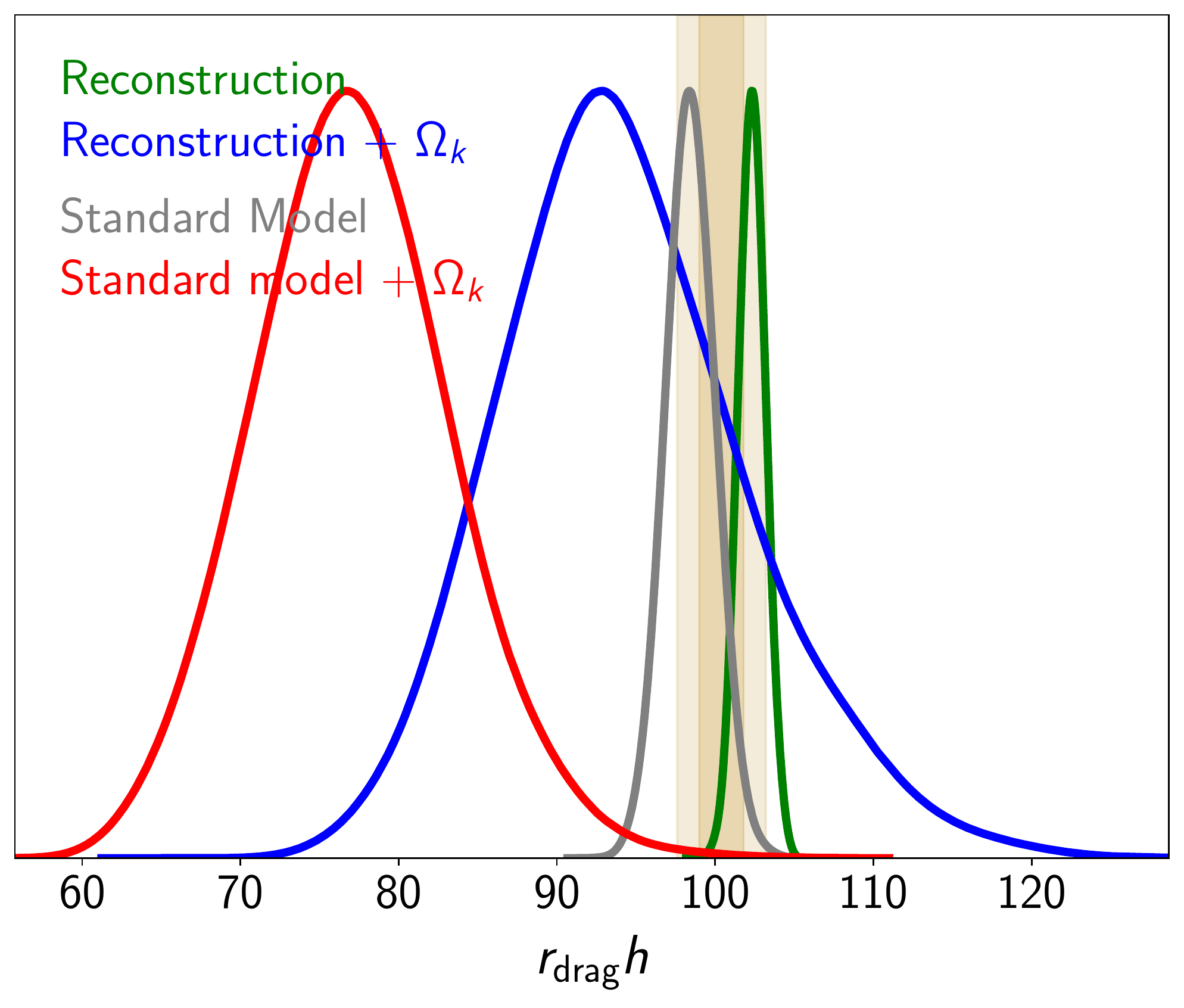}
\caption{\footnotesize\label{fig:Reconrdrag} {\bf Sound horizon:} Estimation of $r_{drag} h$ for our reconstruction in comparison with the results from the {\tt Standard Model} with and without curvature. Results are compared with a model independent estimation of $r_{drag} h$ from supernova and BAO data. $r_{drag} h$ can be used to test consistency of the models with combination of low and high redshift data. } 
\end{figure*}

In~\autoref{fig:Reconrdrag}, we plot the product of $r_{drag}$ (comoving sound horizon at the radiation drag epoch) and $h=H_0/100$. The band is obtained from model independent measurement~\cite{rdragKeeley:2020} using Pantheon Supernovae~\cite{Pantheon:2018} and SDSS eBOSS~\cite{SDSS-IV:2020a,SDSS-IV:2020b} final data release.
{\tt Standard Model +} $\Omega_K$ analysis shows significant tension. When our {\tt Reconstruction} is used as the primordial spectrum, the consistency with the model independent estimate is restored.

\begin{figure*}[!htb]
\centering
\includegraphics[width=0.6\columnwidth]{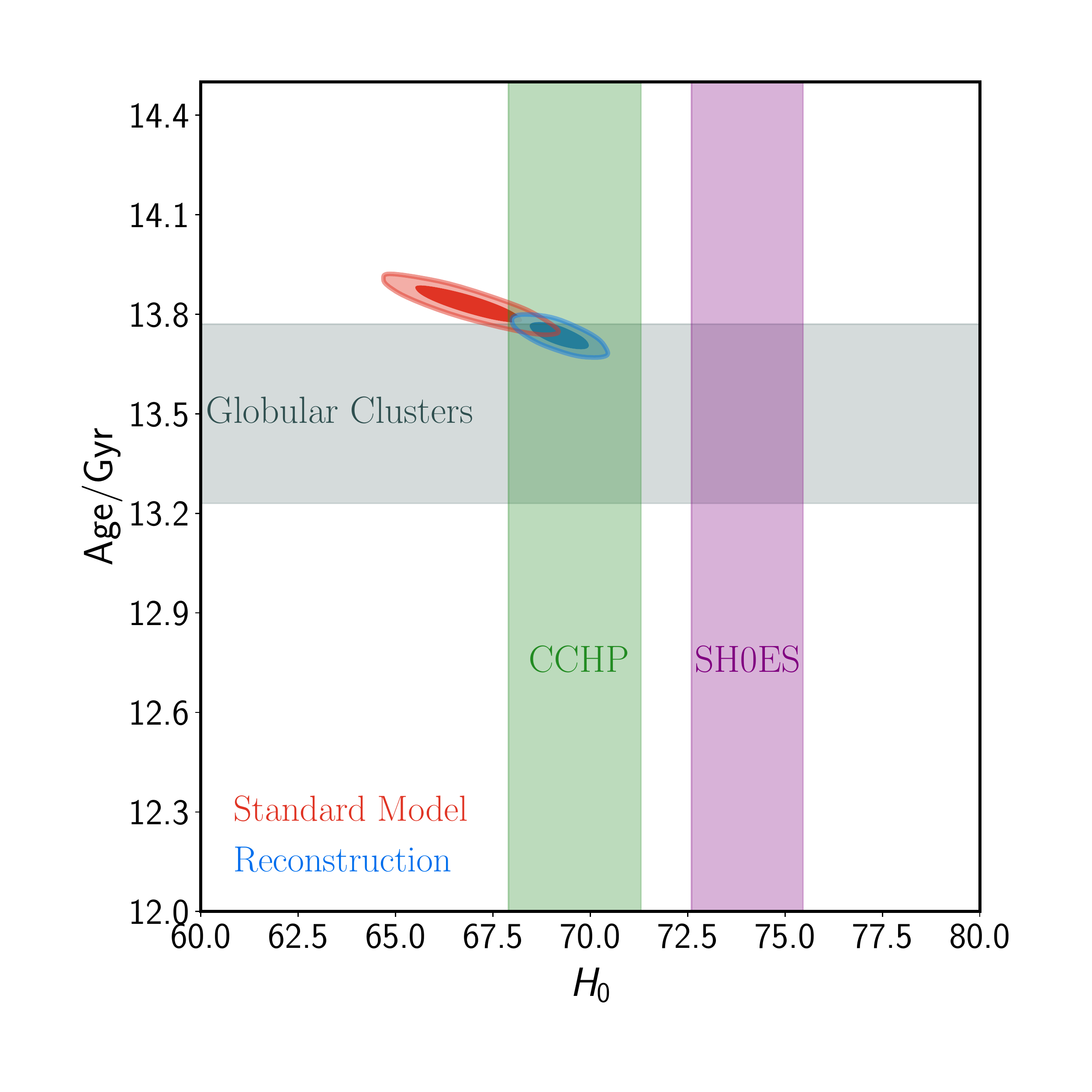}
\caption{\footnotesize\label{fig:ReconAge} {\bf Age of the Universe:} Correlation between Hubble constant and the age of the Universe is plotted as obtained from {\tt Standard Model} and {\tt Reconstruction} analyses. 68\% uncertainty bands are plotted for independent measurement of age from globular clusters~\cite{GlobularClusters:2020,GlobularClusters:2021}, and measurements of Hubble constant from the calibration of the Tip of the Red Giant Branch (CCHP~\cite{CCHP:2020}) and SH0ES (referred here as HST~\cite{Riess:2019}).} 
\end{figure*}

In~\autoref{fig:ReconAge} we plot the correlation between age of the Universe and the Hubble constant obtained in {\tt Standard Model} and {\tt Reconstruction} analysis. Following a recent discussion in~\cite{Age:2021}, we compare the constraints from independent measurements of these two parameters. The bands are plotted corresponding to 68\% uncertainties while the $Age-H_0$ correlations represent 1 and 2$\sigma$ contours. Our reconstructed spectrum shifts the contours towards the local measurements while providing significant improvement in fit to the data \textit{w.r.t} the {\tt Standard Model}. 

From our analyses with the reconstructed spectrum and different extensions to the {\tt Standard Model} it can be summarized that when reconstructed spectrum is used as an initial condition,
\begin{itemize}
    \item we obtain significant improvement in fit to the Planck temperature data, with the extent of fit being nearly identical to the {\tt Standard Model +} $A_{lens}$ and {\tt Standard Model +} $\Omega_{K}$ extension.
    
    \item flat Universe becomes completely consistent with the data.
    
    \item the lensing consistency is satisfied, with $A_{lens}=1$ being preferred.
    
    \item Hubble constant increases significantly compared to the {\tt Standard Model} results. 
    
    \item estimated $\Omega_m-\sigma_8$ or $S_8$ becomes completely consistent with the lensing measurements.
    
    \item $r_{drag}h$ tension obtained in {\tt Standard Model +} $\Omega_{K}$ \textit{w.r.t.} model independent measurement can be completely removed. 
\end{itemize}

 The uncertainty in the determination of the cosmological parameters is less in the analysis with reconstructed spectrum, since here the spectrum is fixed allowing just the variation of amplitude. Since the spectral tilt along with the position and amplitude of the oscillations are fixed, the uncertainty is less than the {\tt Standard Model}. Similarly, due to the fixed form of the spectrum we can not estimate the significance of these oscillations in a Bayesian framework. In the next subsection, where we use analytical template to parametrize the power spectrum, we assess the significance of these oscillations.

\subsection{New Spectrum: analytical template}

In order to estimate the significance of the oscillations we obtain in the reconstruction, parametrization is necessary. On top of the standard power law power spectrum, we add modifications. The modifications come with five extra parameters. A Bayesian analysis reveals which of these parameters can be constrained with the data and are significant. $\mathcal{P}_{New}(k)$ in the equation below represents the template.   
\begin{equation}
    \mathcal{P}_{New}(k) = \mathcal{P}_{Power~Law}(k)\l[1 + \frac{\alpha_1\sin\l(\omega(k-k_0)\r)}{\l(1-\alpha_2\sin\l(\omega(k-k_0)\r)\r)\l(1+\beta(k-k_0)^4\r)}\r]~\label{eq:onespectrum}
\end{equation}

\begin{figure*}[!htb]
\centering
\includegraphics[width=\columnwidth]{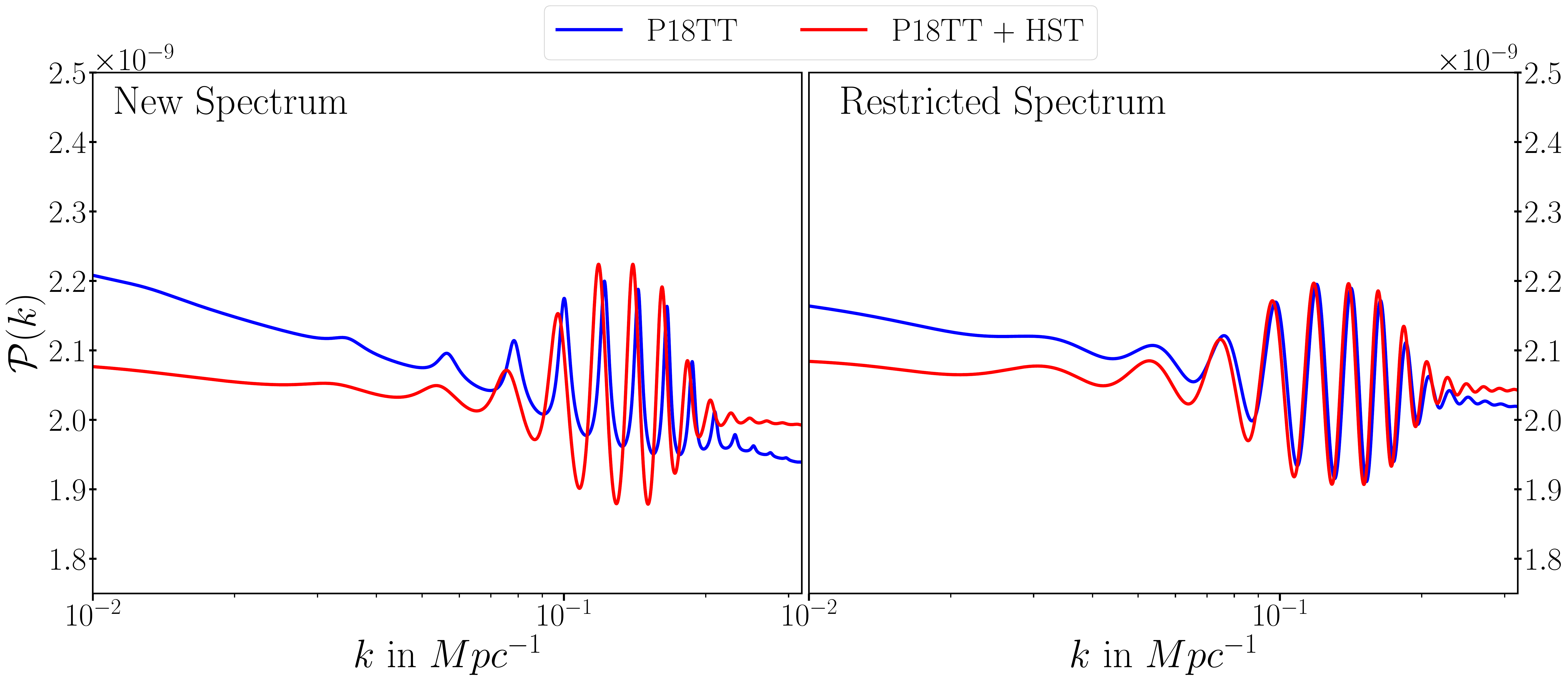}
\caption{\footnotesize\label{fig:BestFitPPS} {\bf Best fit primordial power spectra.} [Left] Best fit {\tt New Spectra} (\autoref{eq:onespectrum}) to the Planck temperature and Planck temperature data with the prior on Hubble constant from~\cite{Riess:2019}. [Right] Best fit {\tt Restricted Spectra}~(\autoref{eq:restrictedspectrum}).} 
\end{figure*}

Here damped sinusoidal oscillations are imposed on the {\tt Standard Model} power spectrum $\mathcal{P}_{power~law}(k)$. The factor $(1-\alpha_2\sin(\omega(k-k_0))$ in the denominator amplifies the peaks compared to the troughs. This template provides similar improvement in fit to the data compared to power law as the reconstructed spectrum provides. Upon comparison with Planck temperature data we do not find more than 1$\sigma$ preference for the parameter $\alpha_2$ (see~\autoref{fig:NewSpectrumTriangle1} in the supplementary information) and therefore the following power spectrum,
\begin{equation}
    \mathcal{P}_{Restricted}(k) = \mathcal{P}_{Power~Law}(k)\l[1 + \frac{\alpha_1\sin\l(\omega(k-k_0)\r)}{1+\beta(k-k_0)^4}\r]~\label{eq:restrictedspectrum}
\end{equation}
with 4 extra parameters provides similar improvement in fit to the Planck temperature data. Since the better fit to the temperature data comes with larger value of Hubble parameter, we compare the {\tt New Spectrum} and {\tt Restricted Spectrum} against the P18TT and P18TT with HST data. In~\autoref{fig:BestFitPPS} we plot the primordial power spectra (both {\tt New Spectrum} and {\tt Restricted Spectrum}), best fit to these two data combinations. Note that when HST data is added, oscillations in the best fit spectra remain localized to similar scales (with marginal shift in phase) compared to the P18TT best fit spectrum. At the same time we notice preference for a higher spectral tilt in the joint analysis.

\subsubsection{Significance and evidence}

 Lensing amplitude extension of the {\tt Standard Model} comes with one extra \textit{unphysical} parameter and the improvement in fit to the data leads to $2.8\sigma$ significance~\cite{Planck:2018param}. Our {\tt New} and {\tt Restricted Spectrum} both come with 4-5 extra parameters compared to the canonical model. Apart from a theoretical support for the spectrum, estimation of statistical significance in proper Bayesian interface is necessary to compare these beyond {\tt Standard Model} scenarios. In~\autoref{tab:evidence} we provide the logarithm of Bayes factor for the extended model compared to the {\tt Standard Model}.
\begin{table}[]
\centering
\begin{tabular}{|l|l|l|}
\hline
Models/Data         & P18TT & \begin{tabular}[c]{@{}l@{}}P18TT + HST\end{tabular} \\ \hline\hline
{\tt New spectrum}        &    -1.14 $\pm$ 0.53    &     2.67 $\pm$ 0.53                                                          \\ \hline
{\tt Restricted spectrum} &  -0.58 $\pm$ 0.52    &  $3.4\pm0.53$                                                                \\ \hline
\end{tabular}
\caption{~\label{tab:evidence} Logarithm of the Bayes factor {\it w.r.t.} to the baseline model. A positive number indicates preference of the model compared to the baseline model.}
\end{table}
For P18TT data, {\tt Restricted spectrum} with 4 extra parameters claims similar significance as the {\tt Standard Model} and the {\tt New Spectrum} remains weakly disfavored. The addition of local measurement of Hubble constant provides \textit{strong evidence} for both the spectra. The {\tt Restricted spectrum} is more favored by the data as it has less parameters with similar improvement in fit. In both the cases we find that the marginalized posterior distribution of the parameter $\alpha_1$ discards the value $\alpha_1=0$ (corresponding to the {\tt Standard Model}) at a confidence level of 99.9\%. 

\begin{figure*}[!htb]
\centering
\includegraphics[width=\columnwidth]{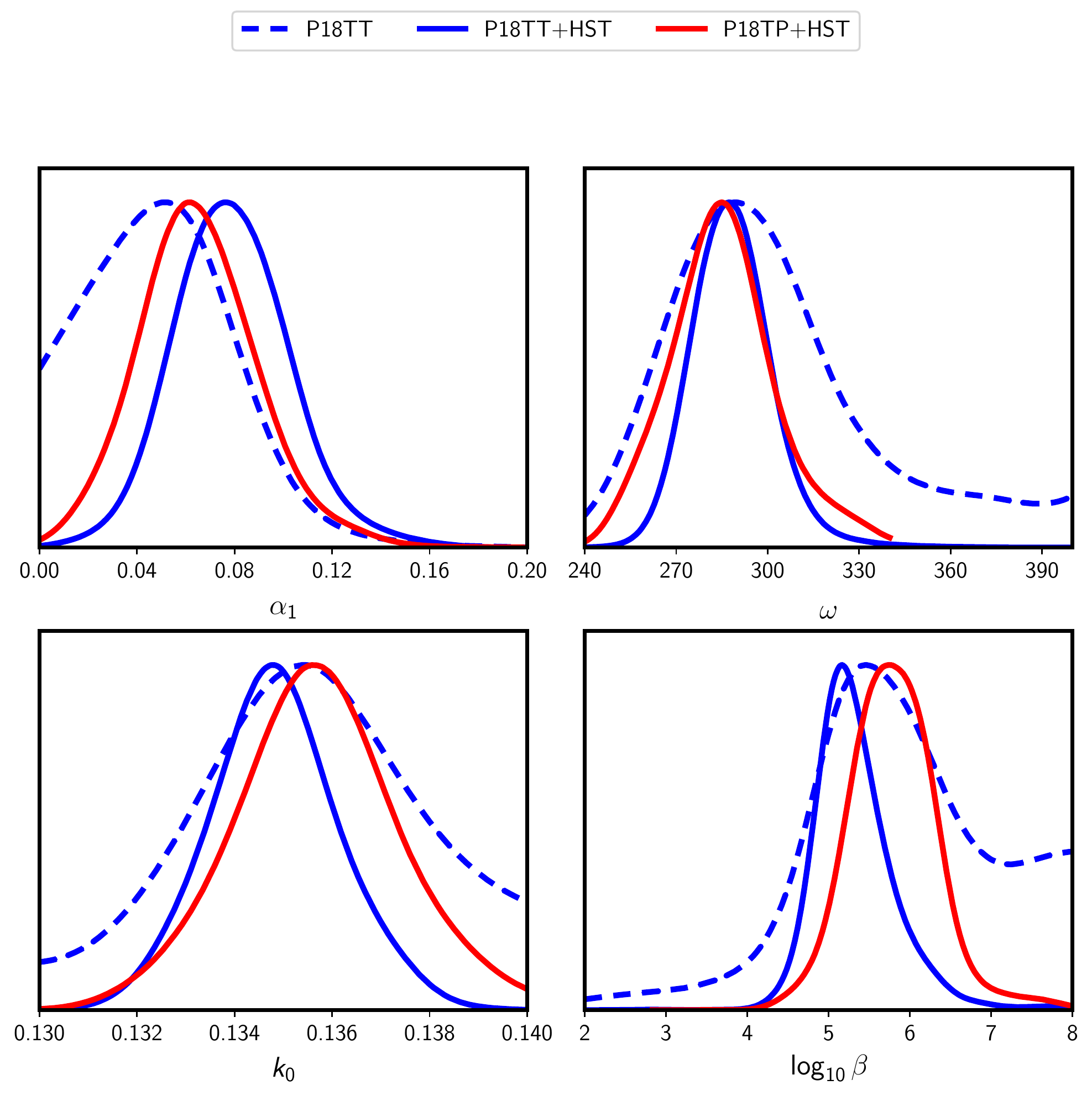}
\caption{\footnotesize\label{fig:NewSpectrumSignificance} {\bf Posterior distributions of parameters in the proposed model.} From Planck temperature data alone we find between 1-2$\sigma$ preference for the  proposed spectra. Use of HST data increases the significance to 99.9\% confidence limit.} 
\end{figure*}

The marginalized posterior distribution for the power spectrum parameters are plotted in~\autoref{fig:NewSpectrumSignificance}.
For the data combination P18TT+HST we find very high significance for the all the parameters except $\alpha_2$ in the {\tt New Spectrum} (see, the triangle plots in~\autoref{fig:NewSpectrumTriangle1} and ~\autoref{fig:NewSpectrumTriangle2}). Both spectra rejects the {\tt Standard Model} (corresponding to $\alpha_1=0$) at 99.9\% confidence limit. When the prior on Hubble constant is not used, there too we see preference (with 82\%-85\% confidence limits) for both {\tt New Spectrum} and {\tt Restricted Spectrum}. Therefore a weakly significant result for these two spectra from Planck temperature data converts to a strong evidence when HST data is used. This evidence does not come in expense to a worse fit to the Planck CMB or HST data. In fact, we find that the $\chi^2$ from Planck data for the best fit obtained with P18TT+HST is better than {\tt Standard Model} best fit $\chi^2$ to P18TT data only. 

Note that the tensions between datasets and anomalies within the CMB data are predominantly driven by the Planck temperature data which is cosmic variance limited. Since signal-to-noise ratio in the polarization data is much lower than temperature data, addition of these data contributes only marginally to these tensions. We therefore only present here the results from Planck temperature and HST. However, it is important to explore whether this significance survives the test of Planck polarization, small scale CMB and other datasets from large scale structure observation. In~\autoref{sec:otherdatasets} we present our analyses with other datasets in detail. There we only analyze the {\tt Restricted Spectrum} as we found the $\alpha_2$ in the {\tt New Spectrum} can not be well constrained by Planck data. 

\subsubsection{Correlations}~\label{sec:correlation}
\begin{figure*}[!htb]
\centering
\includegraphics[width=0.48\columnwidth]{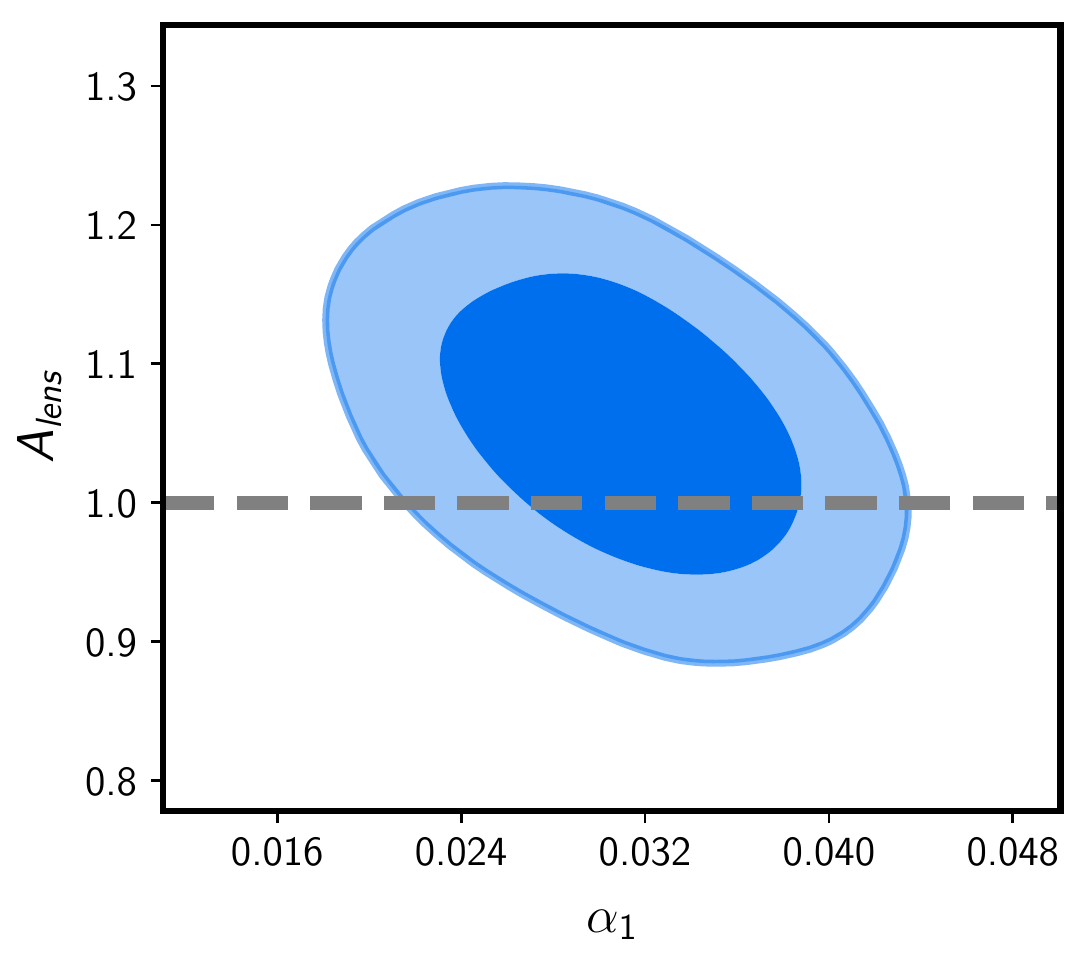}
\includegraphics[width=0.50\columnwidth]{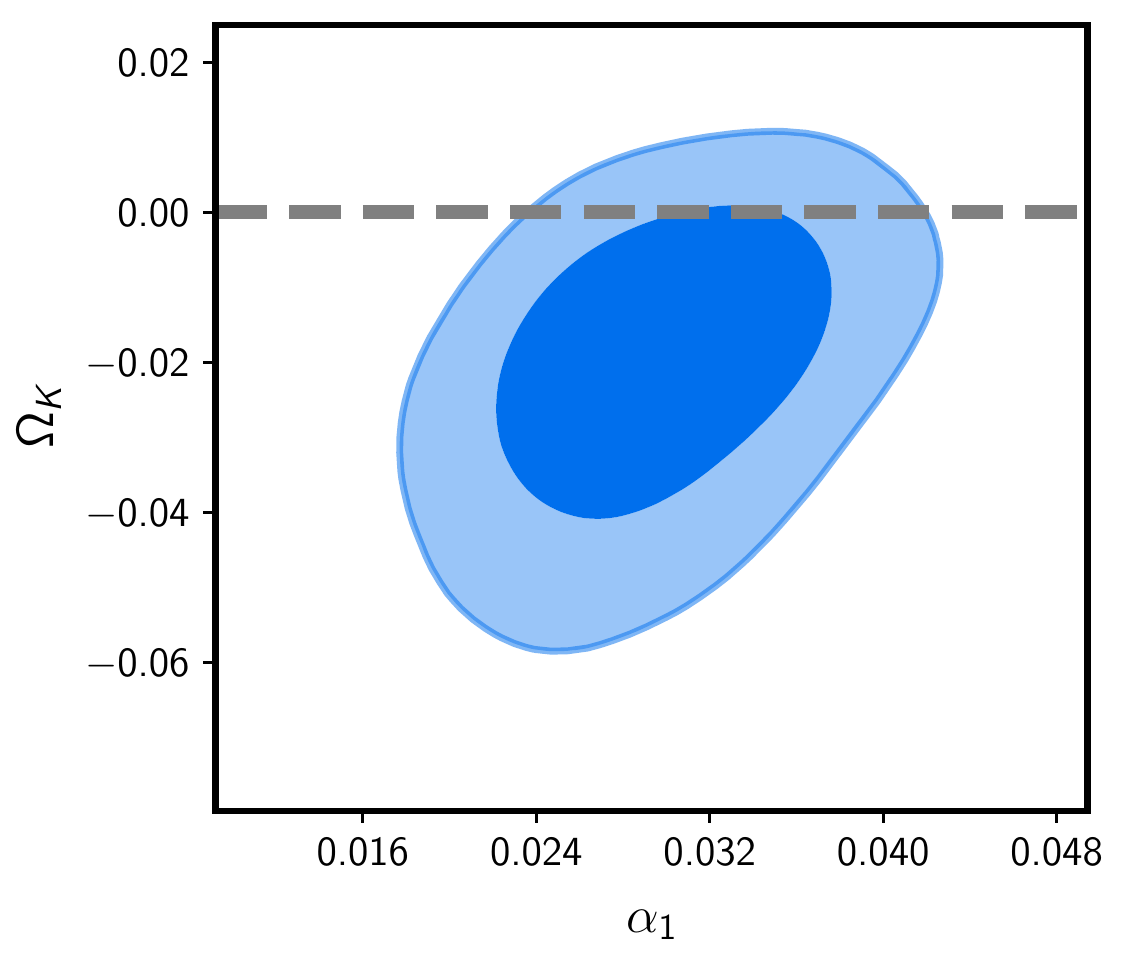}
\includegraphics[width=\columnwidth]{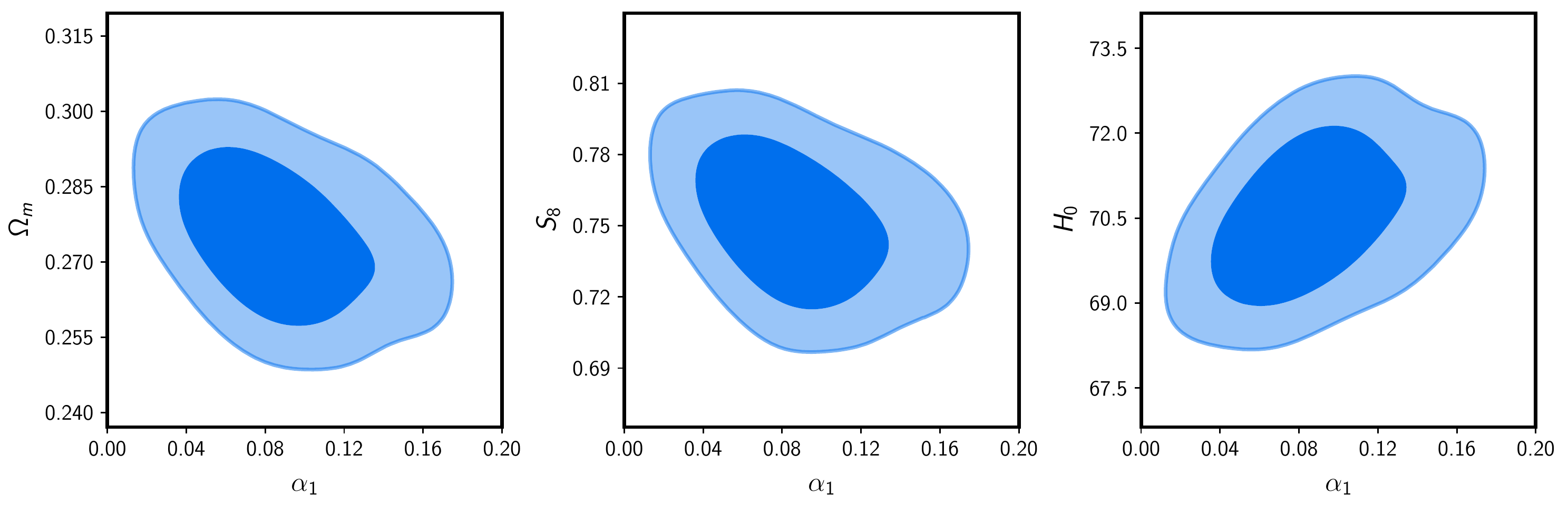}
\caption{\footnotesize\label{fig:Correlation} {\bf Correlations between primordial oscillations amplitude and other parameters:} [Top] Correlation between the amplitude of the linear oscillations ($\alpha_1$) in the {\tt New Spectrum} and the lensing amplitude and the curvature. These correlations are obtained from two separate analyses with Planck temperature data. [Bottom] Correlations between $\alpha_1$ and other cosmological parameters. These correlations are obtained from Planck temperature data and Hubble constant measurement.}
\end{figure*}

In~\autoref{sec:reconstruction} we discussed the solution of closed Universe and lensing anomaly the power spectrum we reconstructed. We also noticed that it removes tensions between different datasets. Since we have been able to parametrize the {\tt Reconstruction} with {\tt New Spectrum}, we can explore the correlations between {\tt New Spectrum} and other cosmological parameters. We perform two analyses with Planck temperature data, namely {\tt New Spectrum + $A_{lens}$} and {\tt New Spectrum + $\Omega_{K}$}. In these two analyses, we fix the nuisance parameters and spectral amplitude ($A_s$), tilt ($n_s$) and the optical depth to their best fit values (obtained from {\tt New Spectrum} analysis with P18TT data). The correlations are plotted in the top panel of~\autoref{fig:Correlation}. Anti-correlation between the amplitude of sinusoidal oscillations ($\alpha_1$) in the {\tt New Spectrum} and $A_{lens}$ and correlation between $\alpha_1$ and $\Omega_K$ is evident.
In the bottom panel of the same figure, we plot the correlation between $\alpha_1$ and $\Omega_m$, $S8$ and $H_0$ found in the analyses of {\tt New Spectrum} with Planck temperature data and HST data. Note that $\alpha_1$ positively correlates with the Hubble parameter that leads to the strong significance for the {\tt New Spectrum}. $\alpha_1$ anti-correlates with matter density and $S_8$. This indicates that higher amplitude of oscillations fits the Planck CMB data with lower matter density and $S_8$.  This makes Planck CMB more consistent with the lensing surveys. Without the prior on the Hubble constant too similar correlations are obtained.

\subsubsection{Extended analysis and robustness of the results}~\label{sec:otherdatasets}
The {\tt New Spectrum} and the {\tt Restricted Spectrum} bring non-standard signature from early Universe at small scales. Therefore such these types of spectra can be tested with small scale ground based CMB observations and also the large scale structure observations. Since this spectrum also shifts the posteriors on background parameters obtained from CMB compared to the {\tt Standard Model} analysis, addition of supernovae distance modulus data and BAO survey datasets can also be used to test the robustness of the obtained results. 

\paragraph{Planck polarization}
Since the tension between the datasets are driven by Planck temperature spectrum (being cosmic variance limited, it is also the most precise data from CMB), we have centered our discussion with Planck temperature only data. Addition of polarization data in the analysis with {\tt Standard Model} provides tighter constraints on the parameters. However when extensions to the model are compared with the data, we find different preferences. For example, $A_{lens}<1$ is preferred by TE+lowE data in contrast to $A_{lens}>1$ preference from TT+lowl, albeit with lower significance compared to temperature data. This internal disagreement leads to a shift in the $A_{lens}$ posterior towards unity. A similar shift towards a $\Omega_K=0$ is also observed in the combined data. In our analysis including Planck polarization data, we find marginal decrease in significance for the amplitude of oscillations in the {\tt New Spectrum}. In~\autoref{fig:NewSpectrumSignificance} we plot the posterior distributions for the parameters in the {\tt Restricted Spectrum}. We find $\alpha_1>0$ at 99.5\% confidence limit for the {\tt Restricted Spectrum} (96\% for the {\tt New Spectrum}). Cosmic variance limited polarization data from future observations are expected to verify the presence of these oscillations in the primordial spectrum to a much better accuracy. 

\paragraph{Extended analysis with datasets from different observations}
Due to possible systematic effects and lack of substantial compatibility (overlap between different datasets leading to double counting of data, sample variance etc.) results from joint analyses with large number of datasets should be treated as consistency checks. Here too we present our extended analyses as such. Using the survey specifications and guidelines to avoid double counting (as discussed in~\autoref{sec:methods}) we compare our model with several data combinations. Different combinations are used to mark the possible contribution of each datasets, \textit{for or against} our proposed model. Here we use the {\tt Restricted Spectrum} for comparison. 

\begin{figure*}[!htb]
\centering
\includegraphics[width=0.49\columnwidth]{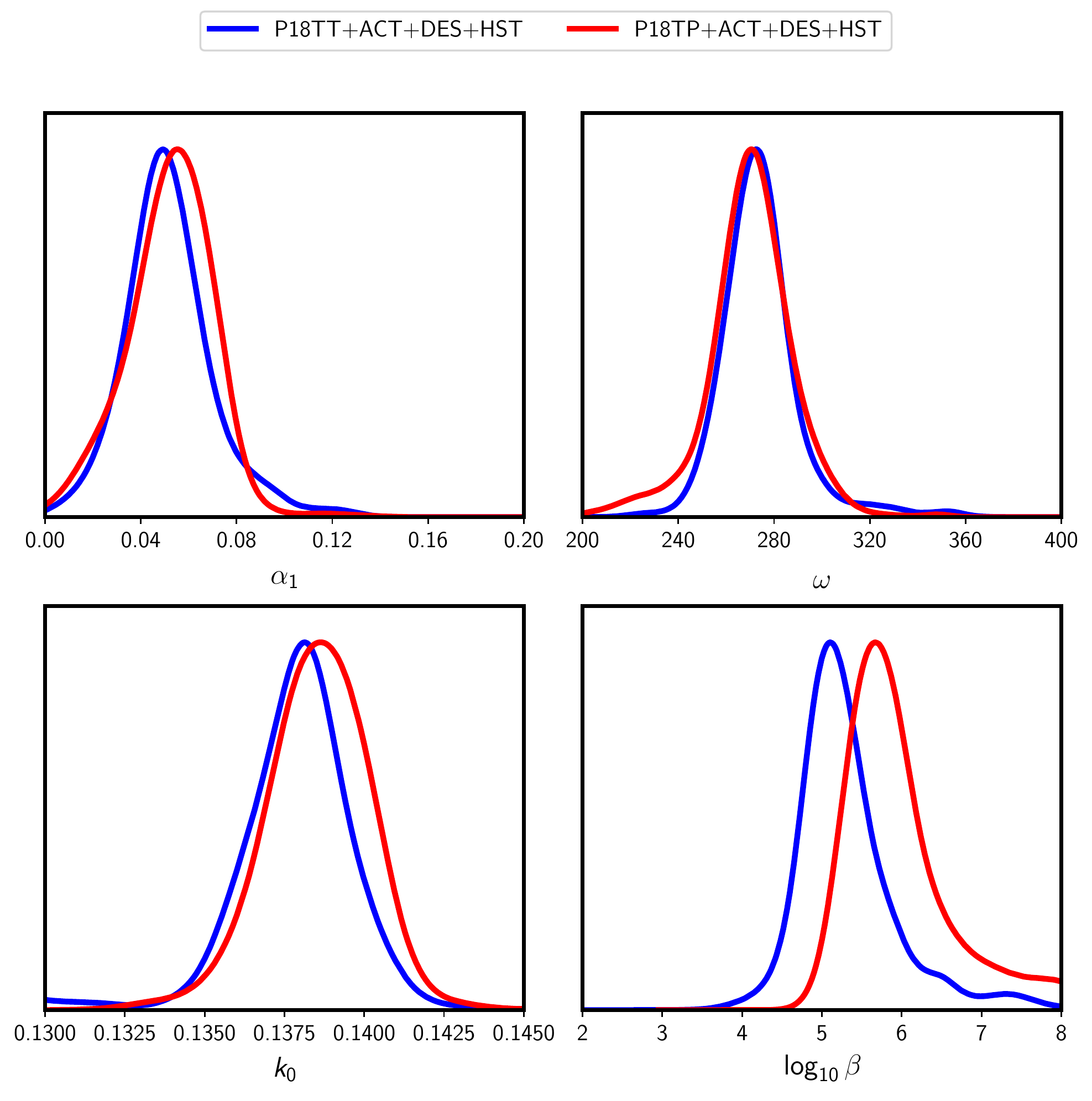}
\includegraphics[width=0.49\columnwidth]{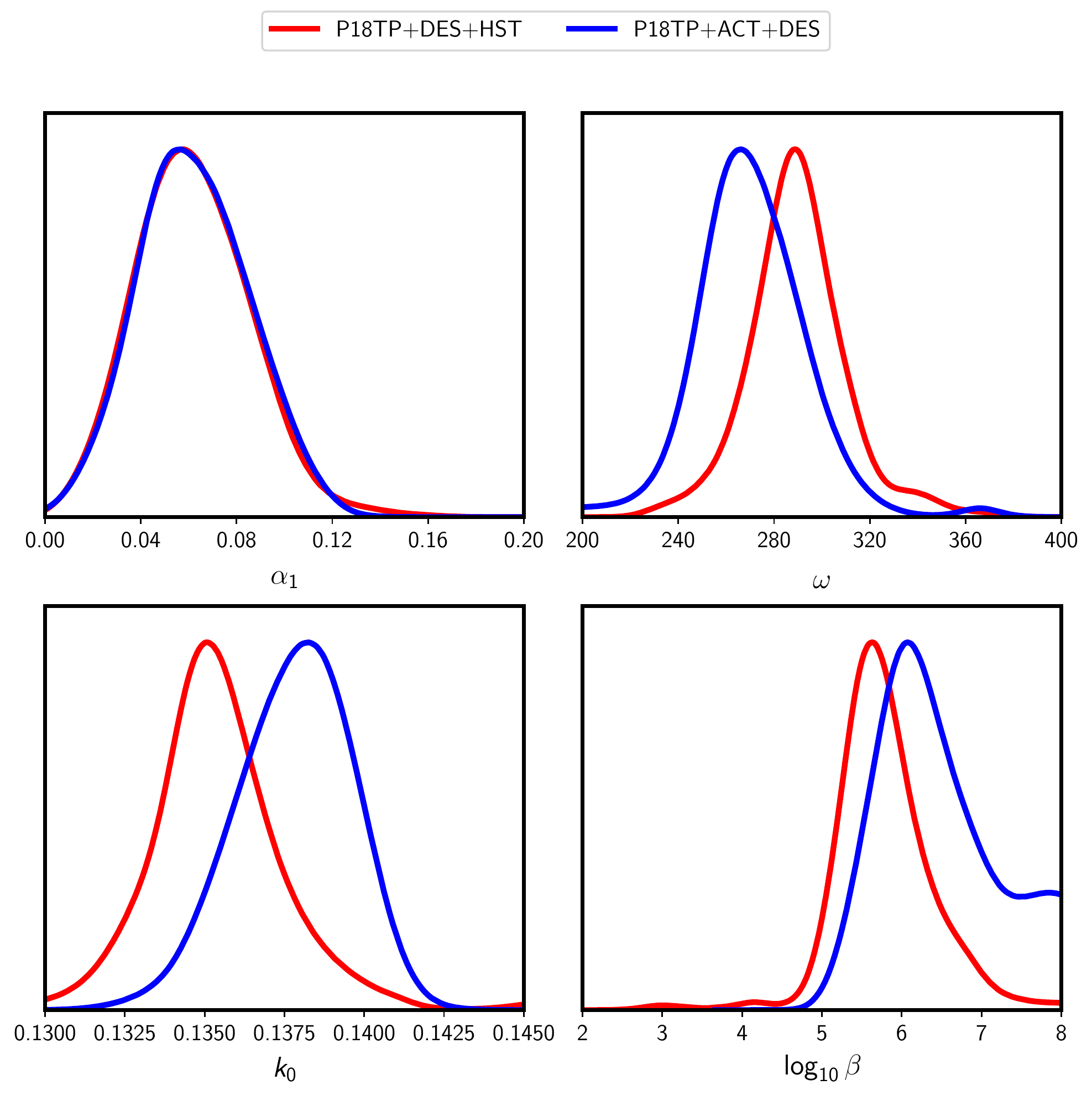}
\caption{\footnotesize\label{fig:ExtendedData}{\bf Constraints from extended joint analyses:} From the extended analysis, the marginalized posteriors of the power spectrum parameters are plotted in two sets, each containing two dataset combinations.}
\end{figure*}

\begin{figure*}[!htb]
\centering
\includegraphics[width=\columnwidth]{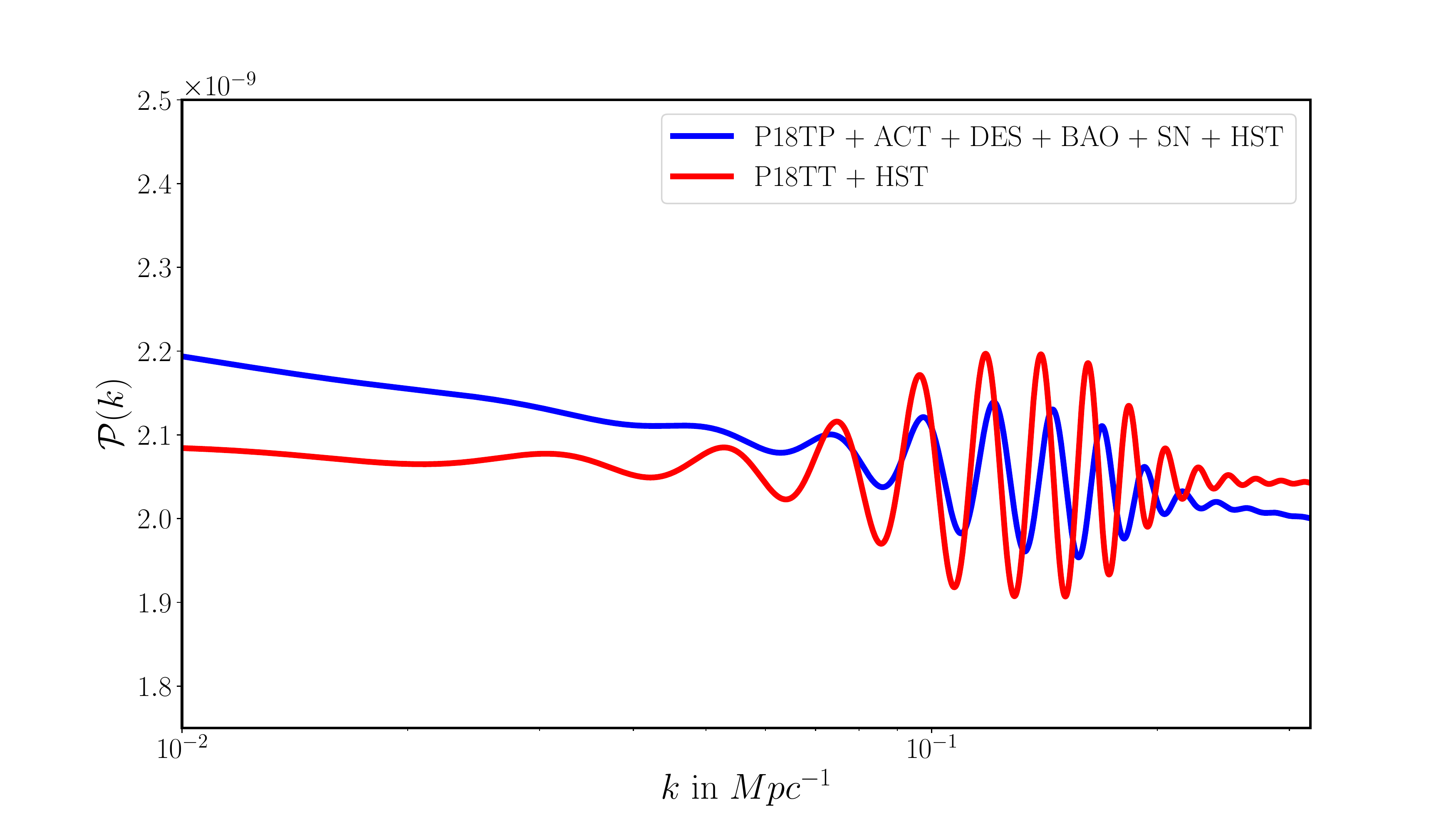}
\caption{\footnotesize\label{fig:BestfitACT}{\bf Best fit spectrum in extended data analysis:} When ACT data is added, the oscillation envelope shifts marginally towards smaller scales. BAO and SN data prefers a higher matter density that correlates negatively with the spectral tilt in fitting CMB data. Therefore in a joint data analysis we find the best fit spectrum to have a redder tilt compared to P18TT+HST best fit.}
\end{figure*}

In~\autoref{fig:ExtendedData}, we plot the marginalized posteriors for the parameters of the {\tt Restricted Spectrum} for four combinations of datasets. In~\autoref{tab:ExtendedDataEvidence}, we list the logarithm of Bayes factor obtained for the {\tt Restricted Spectrum} compared to {\tt Standard Model}. We also list the confidence limit with which $\alpha_1$ rejects the $\alpha_1=0$ ({\tt Standard Model}) that determines the significance of the proposed spectrum. The best fit primordial power spectrum to the \textit{all} data combination (P18TP + ACT + DES + HST + BAO + SN) is plotted in~\autoref{fig:BestfitACT} and compared with P18TT+HST best fit. 

Comparing the tables and plots, we provide the summary of the extended analysis below:
\begin{itemize}
    \item When small scale CMB temperature and polarization data from ACTPol are included in the analysis, the peak position of the oscillation shifts to a smaller scale compared to analyses with Planck CMB and HST.
    \item DES data prefers lower matter density and $\sigma_8$ and therefore smaller $S_8$ compared to {\tt Standard Model} results with Planck. Since our model naturally prefers smaller $S_8$, inclusion of DES data increases the significance compared to the results obtained in P18TP+HST analysis. Correlations between different parameters are plotted in~\autoref{fig:ExtendedDataTriangle1}.
    \item BAO + SN prefers higher matter density. Therefore adding these two datasets drags the parameter space towards {\tt Standard Model} to accommodate higher matter density (see,~\autoref{fig:ExtendedDataTriangle2}).
    \item When DES and ACT data are used for joint analysis with P18TP without the prior on Hubble constant, we still find 99.5\% significance for the proposed model (The marginal likelihood decreases as the decay parameter, $\beta$ is not well constrained).
\end{itemize}

The above findings help to conclude that our model is consistent with extended datasets.

\begin{table}[]
\centering
\begin{tabular}{|l|l|l|}
\hline
Data         & $\ln$[Bayes factor] & Confidence level \\ \hline\hline
\begin{tabular}[c]{@{}l@{}}P18TP + HST \end{tabular} & 1.46 $\pm$ 0.55   & 99.5\% \\ \hline
\begin{tabular}[c]{@{}l@{}}P18TT + ACT + DES + HST \end{tabular} & 2.28 $\pm$ 0.65   & 99.6\%  \\ \hline
\begin{tabular}[c]{@{}l@{}}P18TP + ACT + DES + HST\end{tabular} &  1.94 $\pm$ 0.66    &  98.7\% \\ \hline
\begin{tabular}[c]{@{}l@{}}P18TP + DES + HST \end{tabular}&  2.32 $\pm$ 0.64    &  99.5\%   \\ \hline
\begin{tabular}[c]{@{}l@{}}P18TP + ACT + DES + BAO + SN + HST \end{tabular}&  $-0.34 \pm 0.66$    &  98.5\%   \\ \hline
\begin{tabular}[c]{@{}l@{}}P18TP + ACT + DES \end{tabular}& $-0.85\pm0.66$  &  99.5\%  \\ \hline
\end{tabular}
\caption{\footnotesize\label{tab:ExtendedDataEvidence}{\bf Bayesian evidence and significance.} For different combinations of datasets in our extended analysis we provide the logarithm of Bayes factor \textit{w.r.t.} the {\tt Standard Model}. We also provide the confidence level with which our proposed model rejects the {\tt Standard Model}.}
\end{table}

\subsection{A theory for the spectrum: a potential for inflation}

The primordial power spectrum that we presented here can have different origins, such as inflationary or bouncing Universes. In this article our main results are agnostic about the origin of the oscillations in the spectrum. However, we present a potential for the inflaton (the scalar field $\phi$, that drives the inflation) in~\autoref{eq:Inflation} (similar potential also appeared in~\cite{Antony:2021a}) that follows the primordial oscillations found in the {\tt Reconstruction}. In order to keep the results of this article generic, we leave the analyses with the inflationary potential for another article specific to inflationary models~\cite{Antony:2021b}.

\begin{figure*}[!htb]
\centering
\includegraphics[width=\columnwidth]{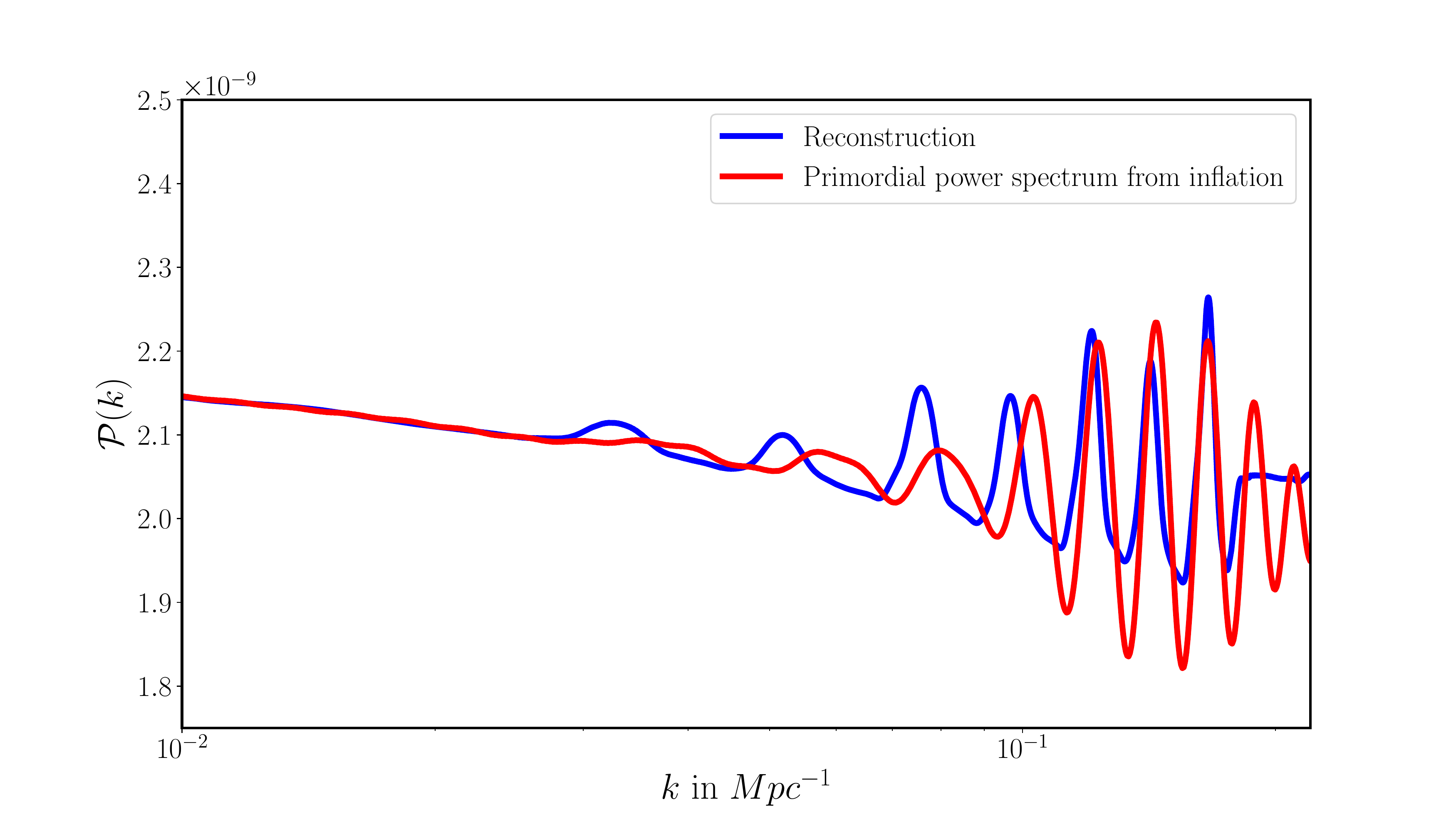}
\caption{\footnotesize\label{fig:Inflation}{\bf A new potential for inflation:} The scalar field potential described in~\autoref{eq:Inflation} generates an inflationary spectrum that closely matches the spectrum obtained with reconstruction.}
\end{figure*}

The potential here is constructed with two parts, firstly a \textit{baseline} slow roll part and a decaying oscillations part. The second part is responsible for the oscillations in the primordial power spectrum. This potential also has four extra parameters but the power spectrum we obtain here depends strongly on the dynamics of the scalar field in this potential. Therefore these parameters do not exactly resemble the parameters in the {\tt New Spectrum} or the {\tt Restricted Spectrum}. We obtain the primordial spectrum by solving the cosmological perturbation equations using BI-spectra and Non-Gaussianity Operator~\cite{BINGO:2013} and we compare it against the {\tt Reconstruction} in~\autoref{fig:Inflation}. The match between these two spectra points out the inflationary model as \textit{a possible theoretical solution for the anomalies and tensions} in cosmology and the appropriate candidate for \textit{new physics}. \\\
\begin{equation}
    V(\phi) = V_{slow~roll}(\phi) + \frac{\alpha \cos\l[\omega(\phi-\phi_0)\r]}{1+\l[\beta(\phi-\phi_0)\r]^2}\label{eq:Inflation}
\end{equation}

\section{Conclusions and outlook}~\label{sec:conclusions}

Our analysis shows a primordial power spectrum with oscillations at small scales and a characteristic frequency of $\omega\sim300$, that we obtain upon reconstruction from CMB data, can resolve the lensing anomaly found in Planck CMB temperature anisotropy data. This power spectrum also removes the strong preference towards a closed Universe and the flat Universe becomes completely consistent with CMB. 

This spectrum also prefers a lower matter density and lower amplitude of linear matter power spectrum that resolves the tension between Planck CMB and a number of lensing and clustering surveys. Importantly, we find the Hubble constant estimated from Planck CMB with this reconstructed spectrum used as the primordial spectrum, shifts to a higher value. This substantially reduces the \textit{Hubble tension} with the SH0ES measurement. The increase in Hubble constant supports a younger Universe which increases the agreement with the independent measurement from globular clusters. 

The reconstructed spectrum can easily be parametrized with decaying linear oscillations. A joint analysis with Planck CMB temperature and prior on Hubble constant indicates strong evidence for the spectrum, with all parameters of the spectrum tightly constrained. Planck temperature and polarization data combined with the Atacama Cosmology Telescope small scale CMB data and the Dark Energy Survey clustering and lensing data are also consistent with the model. Without the priors on the Hubble constant too, our model is supported over the standard model of cosmology. This support for the parametrized spectrum is not prior dominated. At the same time, unlike the modified gravity or dark energy solutions that resolve the tensions in cosmological datasets with significantly enhanced uncertainty, our spectrum only marginally increases the errors in the estimation of cosmological parameters. 

A model of inflation with oscillations in the potential can produce a primordial spectrum similar to the reconstructed spectrum. The potential provides a strong theoretical ground for the spectrum that solves the major anomalies and tensions in the cosmological observations. 

The model we proposed does not rule out the possibility of systematic effects in the tensions between the datasets. However it provides \textit{a new viewpoint} in finding a common solution to the anomalies and tension and thereby greatly enhances its chance to be a suitable tracer of \textit{new physics}.

The early Universe signal we propose in this article is, located at small scales in the CMB data and therefore has complete overlap with ongoing and future large scale structure observations and ground based small scale CMB observations. Correlation function of overdensity exhibits a feature at a comoving radius of about 300 megaparsecs. In future, existence of such type of spectrum can definitely be verified with higher significance.

\section{Methods}~\label{sec:methods}
The methodology adopted in this paper can be divided into parts. Firstly we reconstruct the power spectrum that can mimic the non-standard lensing amplitude signal or the open Universe signal in the CMB temperature anisotropy data. We compare the {\tt Reconstruction} with the variation in lensing amplitude and curvature against the data, to demonstrate how it brings back cosmic concordance. Secondly we parametrize the reconstructed spectrum so that the significance of the spectrum can be estimated.

\subsection{Deconvolution}
CMB angular power spectrum (${\cal C}^{theory}_\ell$) is a convolution between the primordial power spectrum (${\cal P}_k$) of quantum fluctuations and the transfer function. 
\beq
{\cal C}^{theory}_{\ell}=\sum_{k=k_{min}}^{k_{max}}G_{\ell k}{\cal P}_k
\eeq
Given a model of background cosmology, that determines the transfer function ($G_{\ell k}$), the primordial power spectrum can be obtained as a deconvolution from the data. 
This is a matrix inversion problem. Here we use the Richardson-Lucy algorithm~\cite{Richardson:72,Lucy:74} that has been used in the context of CMB with WMAP~\cite{ShafielooIRL:2004,HazraMRLWMAP:2013} and Planck~\cite{HazraMRLPlanck:2014}. 
The power spectrum at any iteration $i+1$ can be obtained from the power spectrum at iteration $i$ following, 
\beq
{\cal P}^{i+1}_k-{\cal P}^{i}_k={\cal P}^{i}_k\times\sum_{\ell=\ell_{min}}^{\ell_{max}}\tilde{G}_{\ell k}\l(\frac{{\cal C}_\ell^{data}-{\cal C}_\ell^{theory(i)}}{{\cal C}_\ell^{theory(i)}}\r)
\eeq
Starting from any initial guess of the power spectrum the method, in each iteration converges towards a solution of maximum likelihood. Usually ${\cal C}_\ell^{data}$ is used in the deconvolution process. While the use of the data points out the anomalies after smoothing, it also brings in certain amount of noise in the reconstructed primordial power spectrum. Since here we reconstruct the primordial spectrum that mimics excess lensing, we treat the best fit {\tt Standard Model + $A_{lens}$} best fit to the P18TT dataset as the data. Treating this smooth angular power spectrum as the data guarantees modifications to the primordial power spectrum only within the scales where lensing is relevant. The possibility of mimicking lensing effect with the primordial power spectrum was first discussed in~\cite{HazraMRLWMAP:2013}. Two of the authors of this article had also implemented this method as a solution to the $H_0$ tension in~\cite{HazraHST:2018}.  

In~\cite{DiValentino:2019} the degeneracy between curved Universe and the non-standard lensing amplitude in the CMB temperature power spectrum was discussed in detail. A primordial spectrum that mimics the lensing amplitude in the CMB temperature spectrum is therefore expect to mimic the signal from the closed Universe as well. Here the crucial step is to fix the background parameter for the reconstruction, in order to get the transfer function. The background parameters corresponding to the best fit closed Universe scenario is not helpful as the reconstructed power spectrum will be consistent to the CMB data with a very low value of Hubble constant. On the other hand, since {\tt Standard Model + $A_{lens}$} best fit prefers a Hubble constant closer to the local measurement, we use the background parameters from the {\tt Standard Model + $A_{lens}$} best fit to P18TT data.~\footnote{Note that while for reconstruction we use these values, we vary all the background parameters while we compare the reconstructed spectrum with the data.}

\begin{figure*}[!htb]
\centering
\includegraphics[width=\columnwidth]{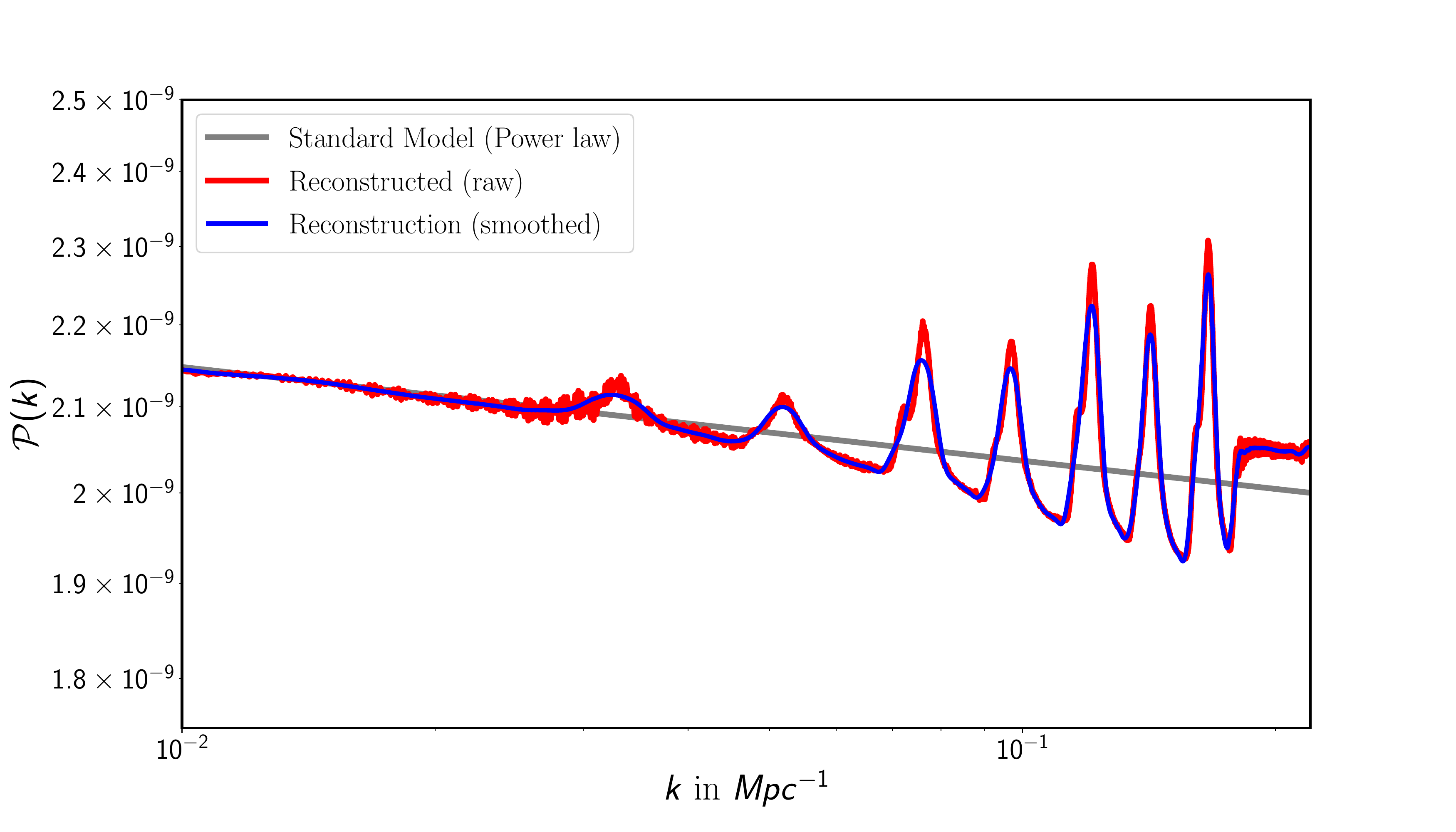}
\caption{\footnotesize\label{fig:ReconSmooth}{\bf Reconstructed spectrum.} The primordial power spectrum obtained through Richardson-Lucy deconvolution that mimics the excess lensing signal in the CMB temperature anisotropy spectrum. We also plot the spectrum smoothed with Savitzky-Golay filter to highlight the main features in the spectrum.}
\end{figure*}

\begin{figure*}[!htb]
\centering
\includegraphics[width=\columnwidth]{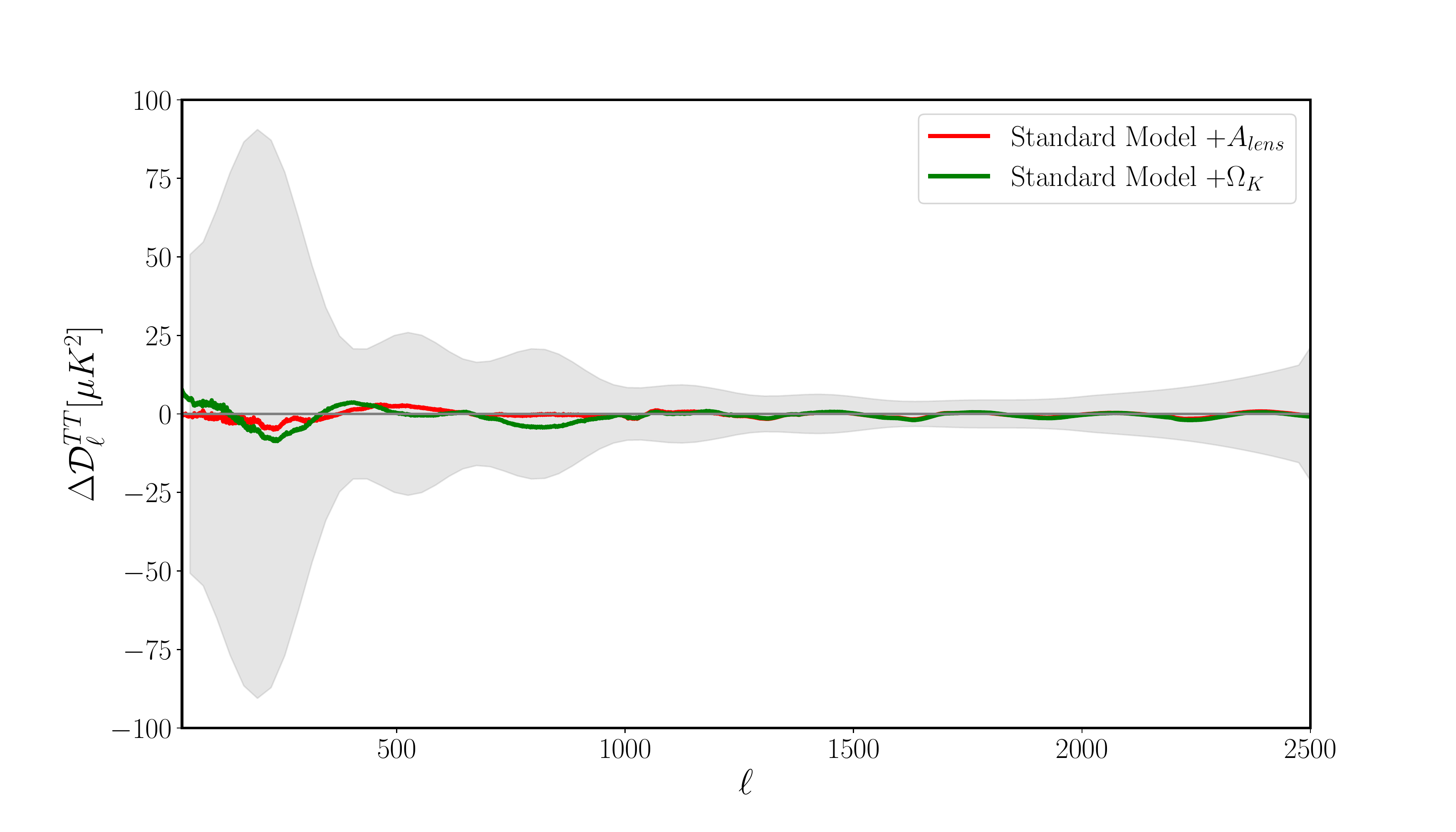}
\caption{\footnotesize\label{fig:Recon-Res-Alens-Omk}{\bf Degeneracy:} Differences between the angular power spectra from {\tt Reconstruction} and {\tt Standard Model + $A_{lens}$} and between {\tt Reconstruction} and {\tt Standard Model + $\Omega_{K}$} are plotted. The Planck error band (68\%) is also provided in grey. Note that the differences are negligible \textit{w.r.t.} the uncertainties. It clearly shows complete degeneracy between these three different physical scenarios.}
\end{figure*}

The deconvolution needs unlensed CMB spectrum. We compute the lensing contribution with standard $A_{lens}=1$ (for the {\tt Standard Model + $A_{lens}$} best fit values of the 4 background and 2 power spectrum parameters). From the {\tt Standard Model + $A_{lens}$} best fit lensed angular power spectrum, we subtract this lensing contribution computed for $A_{lens}=1$ and obtain \textit{unlensed} angular spectrum. This \textit{unlensed} spectrum contains the excess lensing signature due to $A_{lens}>1$. Using this unlensed angular power spectrum, when we reconstruct the primordial power spectrum, it is expected that with $A_{lens}=1$, but assuming the reconstructed spectrum as the primordial spectrum, we recover {\tt Standard Model + $A_{lens}$} best fit angular spectrum (for the same background parameters).  
In~\autoref{fig:ReconSmooth} we plot the reconstructed spectrum in red. As expected, at large scales the spectrum follows power law and at intermediate and small scales, where lensing is significant, we notice certain oscillations. At scales smaller than $0.2~Mpc^{-1}$, the spectrum again takes a power law form as we use the angular spectrum till multipole 2500 (smallest scale probed by Planck) for the reconstruction.   
We find the angular power spectrum from the reconstructed primordial spectrum is nearly identical to the {\tt Standard Model + $A_{lens}$} best fit and also to the {\tt Standard Model + $\Omega_{K}$} best fit. We then compare our {\tt Reconstruction} with P18TT data allowing background and nuisance parameters to vary. We also allow a variation in overall amplitude of the primordial reconstructed spectrum. In~\autoref{fig:Recon-Res-Alens-Omk} we plot the differences in the best fit angular power spectra obtained from {\tt Reconstruction} and {\tt Standard Model + $A_{lens}$} analysis. We also plot the difference \textit{w.r.t} the {\tt Standard Model + $\Omega_{K}$} best fit. The shaded region marks the 68\% uncertainty band in the binned Planck temperature data. In comparison to the error band, the differences are negligible, that demonstrates the degeneracy.
\subsection{Getting the analytical template}
\begin{figure*}[!htb]
\centering
\includegraphics[width=\columnwidth]{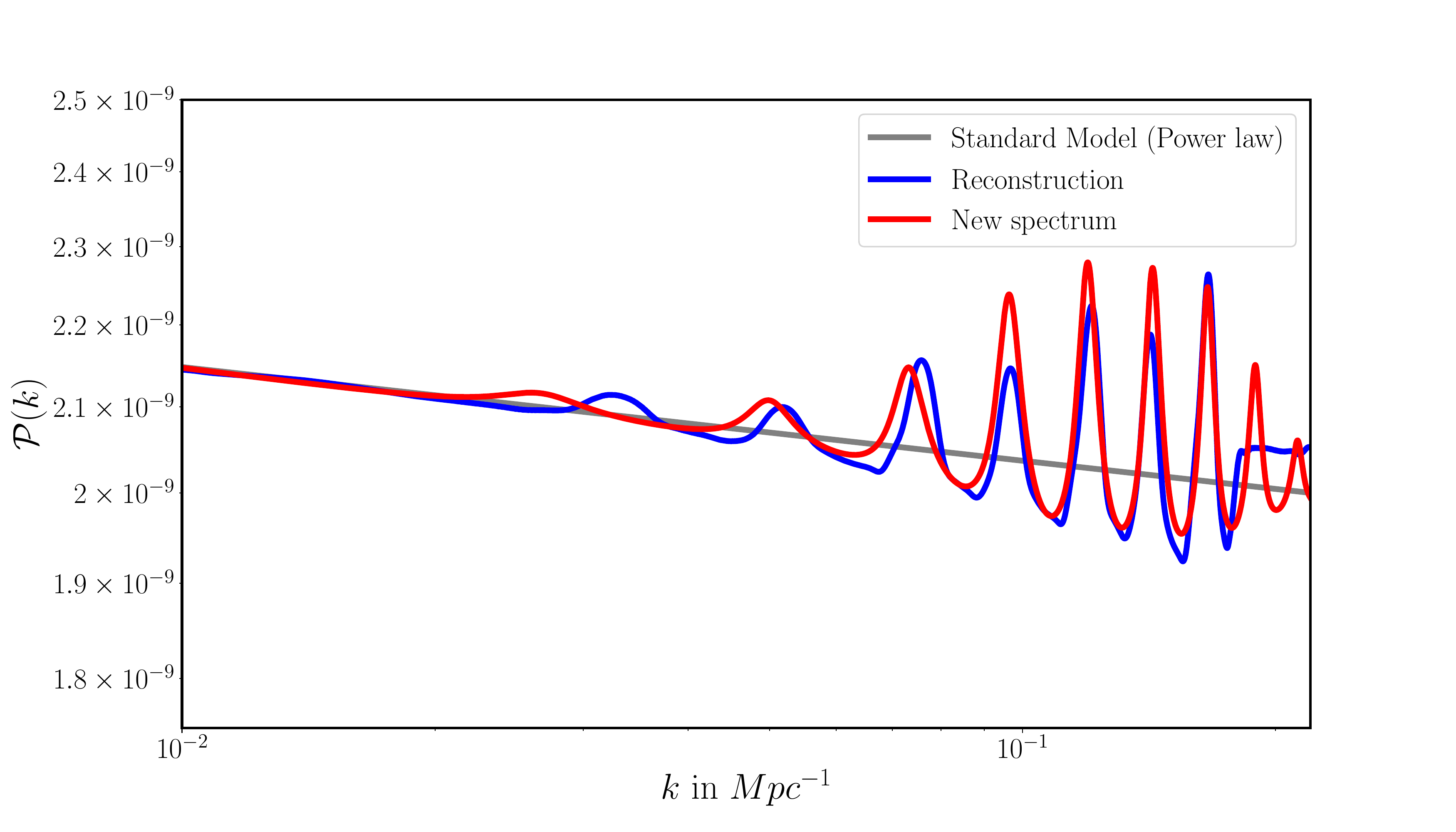}
\caption{\footnotesize\label{fig:Newspec_match_recon}{\bf Comparison between {\tt Reconstruction} and the {\tt New Spectrum}.} The analytical function in~\autoref{eq:onespectrum} is plotted against the reconstructed power spectrum. Note that the template is able to match the major features in the reconstructed spectrum.}
\end{figure*}

Following the analysis with the reconstructed spectrum, in order to compute the significance of the reconstructed spectrum, parametrization is necessary. We identify three major characteristics in our {\tt Reconstruction},
\begin{itemize}
    \item Oscillations in the spectrum are linear in $k$
    \item Oscillations decay towards large scales
    \item Peaks are more pronounced than the troughs 
\end{itemize}
Using these information, we parametrize our {\tt Reconstruction} using~\autoref{eq:onespectrum}. A comparison between the {\tt Reconstruction} and its analytical match is plotted in~\autoref{fig:Newspec_match_recon}. We do not perform a statistical comparison between these two spectra as ultimately our goal is to fit the data which can allow certain variation in the analytical spectrum \textit{w.r.t} the {\tt Reconstruction}. This function has five extra parameters compared to the power law spectrum. However not all the parameters are essential in obtaining a better fit to the data and their significance can be obtained from the posterior distribution of each parameter.

\subsection{Essentials on parameter estimation}
We would like to inform the readers that our main results are based on Planck data and in some cases the HST data. Analyses with the use of other datasets should be viewed as tests of consistency.

\subsubsection{Likelihoods}
 In~\autoref{tab:data} we list the data and likelihoods along with their acronyms that we use in this article.
\begin{table}[!htb]
\begin{tabular}{|l|l|l|}
\hline
Acronym & Data                              & Comments                                                                                                                     \\ \hline
P18TT   & TT+lowl+lowE (CMB)                & \begin{tabular}[c]{@{}l@{}}Planck CMB temperature anisotropy \\ and large scale polarization anisotropy~\cite{Planck:2019Like}\end{tabular}         \\ \hline
P18TP   & TTTEEE+lowl+lowE (CMB)            & \begin{tabular}[c]{@{}l@{}}Planck CMB temperature \\ and polarization anisotropy~\cite{Planck:2019Like}\end{tabular}                                \\ \hline
DES     & Clustering + Weak lensing & \begin{tabular}[c]{@{}l@{}}Dark Energy Survey year 1 likelihoods\\  on galaxy clustering and weak lensing~\cite{DES:2017}\end{tabular}       \\ \hline
ACT     & ACT-TTTEEE (CMB)                  & \begin{tabular}[c]{@{}l@{}}Atacama Cosmology Telescope DR4 CMB \\ likelihood on temperature and polarization~\cite{ACT:2020}\end{tabular}    \\ \hline
BAO     & Baryon Acoustic Oscillation       & \begin{tabular}[c]{@{}l@{}}Distance ratio measurements from 6df \\ galaxy survey, Sloan Digital Sky Survey~\cite{BAO6DF:2011,BAODR7:2014,BAODR12BOSS:2016}\end{tabular}      \\ \hline
SN      & Pantheon18 Supernovae             & \begin{tabular}[c]{@{}l@{}}Distance modulus from 1048 \\ Supernovae samples~\cite{Pantheon:2018}\end{tabular}                                     \\ \hline
HST     & $H_0$ local measurement (SH0ES)           & \begin{tabular}[c]{@{}l@{}}Local measurement of Hubble constant from~\cite{Riess:2019}\end{tabular} \\ \hline
\end{tabular}
\caption{~\label{tab:data} {\bf Data used in this article.} We list the data that we use for our analyses. We have used the same acronyms throughout this article.}
\end{table}

\paragraph{Planck:} We use Planck CMB likelihood~\cite{Planck:2019Like} of temperature anisotropy mainly, along with low-$\ell$ E-mode polarization likelihood. In some cases of extended data analysis, we also use Planck polarization likelihoods. We use {\tt PlikHM} binned likelihood from the official release for the high multipoles. For low multipoles ($\ell=2-29$) we use {\tt commander} based likelihood (lowl) for temperature anisotropy. Low-$\ell$ E-model polarization likelihood (lowE) is based on the cross spectrum from the high frequency instrument. We allow all the nuisance parameters to vary in all the analyses except for one where we explore the correlation between the amplitude of the oscillation in the primordial spectrum and the lensing amplitude of curvature discussed in~\autoref{sec:correlation}. We use the recommended priors on the foreground and calibration parameters. 

\paragraph{ACTPol:} The Atacama Cosmology Telescope recently released their DR4 data and likelihood in~\cite{ACT:2020}. We use both the auto and cross correlation in temperature and polarization data from ACTPol. The minimum multipole bin provided in temperature spectrum  is $\ell_{min}=600$. For EE and TE, $\ell_{min}=350$. $\ell_{max}=4125$ is used as the maximum multipole. We use these specifications when we use ACTPol data alone. In that analysis we also use a prior on the reionization optical depth $\tau=0.065\pm0.015$. Note that this value optical depth is higher than the Planck 2018 estimation of $\tau=0.054\pm0.007$. However, to be consistent with ACTPol analyses in~\cite{ACT:2020,HillACT:2021} we use this prior on optical depth in the ACTPol-\textit{only} analysis. While we find the similar frequency of oscillations (corresponding to P18TT best fit) favored by the ACT data (with a $\Delta\chi^2\sim8$ compared to power law) as well, since ACT data alone prefers a spectral tilt greater than unity, we do not present the ACTPol-\textit{only} analysis results. Most of our ACTPol analyses are performed therefore with Planck data where we use low-$\ell$ polarization likelihood from Planck. Note that the overlap with Planck temperature data with ACT at intermediate and small scales requires a cutoff in the multipole ranges. In such cases, following the ACTPol-collaboration analysis we discard the ACT TT data before $\ell=1800$. In our Planck and ACT combined analyses, we fix the ACTPol nuisance parameter $y_p$ (overall polarization efficiency) to 1.

\paragraph{DES:} 
Galaxy clustering and weak lensing data from the Dark Energy Survey~\cite{DES:2017} year 1 (DES Y1) have been used in some of our analyses. This survey results help in providing a constraint on the matter density and the $\sigma_8$ or the $S_8$ parameter. We use the combined dataset using galaxy clustering, galaxy-galaxy lensing and cosmic shear. We allow all the nuisance parameters to vary.

\paragraph{Baryon Acoustic Oscillations:} 
The BAO dataset that we use here is composed of several surveys. At $z=0.15$, we use the distance measurement obtained from the Main Galaxy Sample~\cite{BAODR7:2014}. At $z=0.106$ we use the sound horizon (at drag epoch) and distance ratio measurement from 6dF Galaxy Survey~\cite{BAO6DF:2011}. We also include the BAO datasets from final galaxy clustering data (termed as `consensus sample') with the Baryon Oscillation Spectroscopic Survey (BOSS) at $z=0.38,0.57$ and 0.61~\cite{BAODR12BOSS:2016}.

\paragraph{Supernovae:} 
We use type Ia Supernovae from the `Pantheon' compilation~\cite{Pantheon:2018}. It consists of 1048 SN samples spanning between redshifts 0.01 and 2.3. We use SN data always in combination with BAO datasets. The use of SN and BAO data in the analysis is mainly to provide constraints on matter density alongside other datasets.  

\paragraph{HST:} In some analyses, to estimate the significance of the {\tt New Spectrum} we use the constraint on Hubble parameter, $H_0=74.03\pm1.42$ from~\cite{Riess:2019}.

\subsubsection{Sampling}
We use Code for Anisotropies in the Microwave Background ({\tt CAMB})~\cite{CAMB} to compute the CMB and matter power spectrum. It is also used to calculate distance modulus, distance ratios that are compared against specific observational data. Markov Chain Monte Carlo (using {\tt CosmoMC})~\cite{CosmoMC} is used for the analysis with our {\tt Reconstruction}. In order to estimate the significance of the analytical template proposed as the {\tt New Spectrum}, we perform nested sampling using {\tt Polychord}~\cite{Handley:CC}. We compare the marginal likelihoods of the proposed model with the {\tt Standard Model} ensuring the common parameters in both models have the same priors. 

\subsubsection{Priors}
We use flat priors on background parameters and parameters in the primordial spectrum as tabulated in~\autoref{tab:PriorRanges}. When we use nested sampling using {\tt Polychord} add-on of {\tt CosmoMC}, we use smaller priors on background parameters and the overall amplitude and tilt of the primordial spectrum, ensuring the posteriors stay well inside the priors. For {\tt Restricted Spectrum} we keep the same prior ranges apart from fixing $\alpha_2=0$. 
Upper limit of $\alpha_1$ is chosen such a way that it produces sufficiently large oscillations that would be ruled out by P18TT only. The upper limit of the parameter $\alpha_2$ ensures that the spectrum remains positive definite for all choices of other parameters. The frequency $\omega$ is allowed to vary within a wide range around the frequency that matches the {\tt Reconstruction}. Since the decay parameter $\beta$ is allowed to vary beyond an order of magnitude, we use uniform prior on $\log_{10}\beta$. For the upper limit of $\beta$ the power spectrum only contains one peak (oscillations decay very fast) and for the lower limit, the spectrum has sinusoidal oscillations ranging from largest to the smallest scales probed by Planck. Since our spectrum has sinusoidal oscillations, unless we have very large $\beta$, a very wide prior on $k_0$ will result in repeating patterns. Therefore we allow $k_0$ within half a period of oscillation on either side of mean $k_0$ that matches the reconstruction (see ~\autoref{fig:Newspec_match_recon} for the comparison). When ACT data is used, we increase the upper limit on $k_0$ since ACT probes smaller scales compared to Planck. Priors on the nuisance parameters are used following the survey guidelines. 
\begin{table}[]
\centering
 \begin{tabular}{| l | l | l | }
 \hline
Model	&	Parameter	&	Prior range\\
\hline		
Common parameters	&	$\Omega_\mathrm{b}h^2$             	&	$[0.005,0.1]$	\\
	&	$\Omega_\mathrm{c}h^2$    	&	$[0.001,0.99]$	\\
	&	$100\theta$ 	&	$[0.5,10]$	\\
	&	$\tau$     	&	$[0.01,0.8]$	\\
	&	$\ln[10^{10}A_s]$ 	&	$[1.61,3.91]$	\\
	&	$n_s$     	&	$[0.01,0.8]$	\\
\hline
{\tt New Spectrum}	&	$\alpha_1$	&	0-0.2		\\
	&	$\alpha_2$	&	0-0.9\\
	&	$\omega$	&	240-400\\
	&	$k_0$	&	0.13-0.14	\\
	&	$\log_{10}\beta$	&	2-8	\\
	\hline

\end{tabular}
\caption{~\label{tab:PriorRanges}{\bf Uniform prior ranges.} Uniform priors used for the background and power spectrum parameters in the MCMC analyses.}
\end{table}

\subsubsection{Settings}
For {\tt Standard Model} and its extensions, we use the default specifications in {\tt CAMB}. By default, the angular power spectrum is calculated for a few multipoles and then interpolated in between. In our analyses with primordial power spectrum containing oscillations ({\tt Reconstruction, New Spectrum} and {\tt Restricted Spectrum}), we compute CMB angular power spectrum at every multipole except for two cases. The transfer function computation becomes significantly time consuming in non-flat Universe. In such cases, we use interpolation in the angular power spectrum. Secondly, for ACT data, we calculate the angular spectra till $\ell=6000$. Here too computation of transfer function is extremely time consuming. Therefore, here also we use interpolation. However, in these two cases, we calculate the spectrum for more number of points compared to the default configuration and we make sure that the difference in \textit{goodness of fit} to the data is negligible between our choice and no-interpolation scenarios. We use non-linear lensing in obtaining the lensing contributions to the angular power spectrum. We stop our Markov Chain Monte Carlo when Gelman-Rubin convergence parameter ($R-1$) reaches 0.015. In nested sampling analyses with {\tt PolyChord}, we use 512 live points.

\paragraph{Acknowledgements}The authors acknowledge the use of computational resources at the Institute of Mathematical Science’s High Performance Computing facility [Nandadevi].

\bibliography{main}

\appendix

\section{Supplementary materials - I: Implications for the large scale structure}~\label{appendix:lss}

\begin{figure*}[!htb]
\centering
\includegraphics[width=\columnwidth]{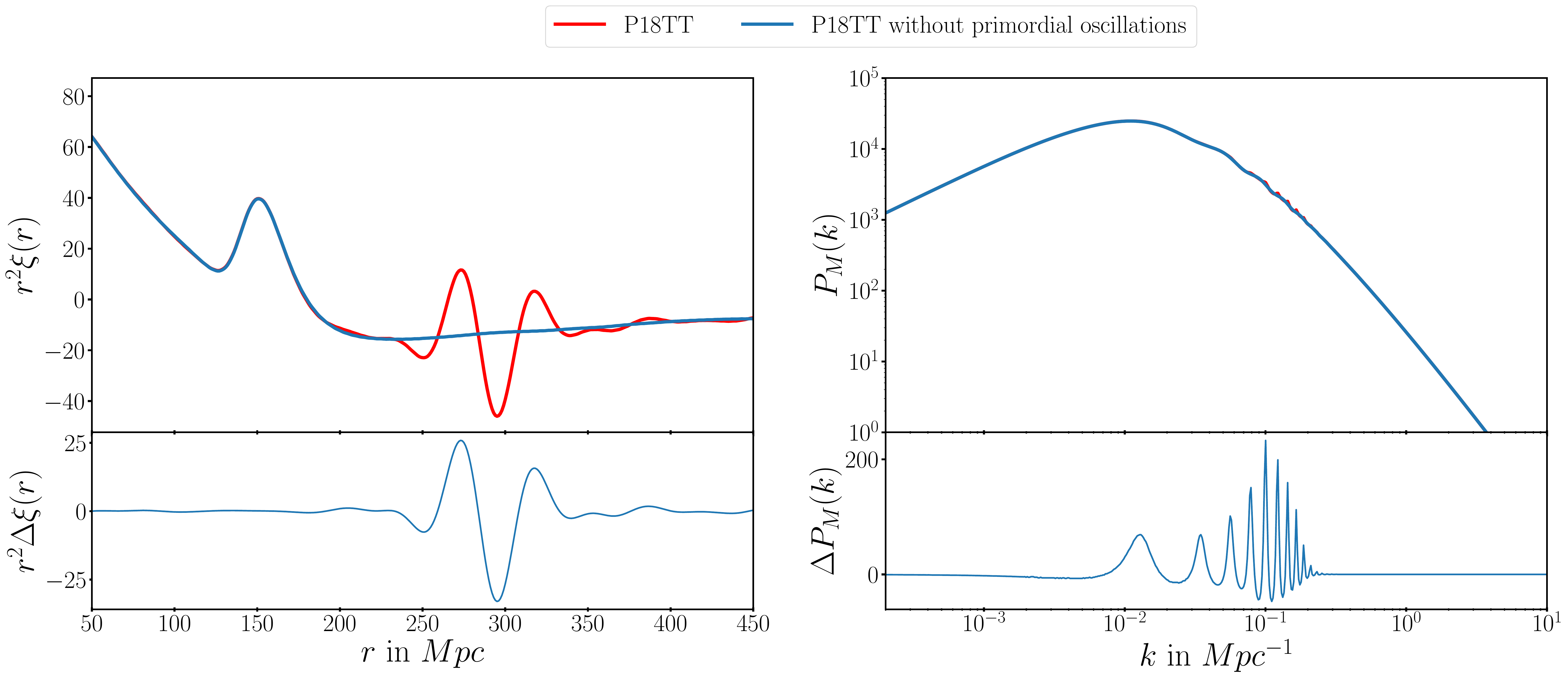}
\caption{\footnotesize\label{fig:CorrelationMatterPower} {\bf Correlation function and matter power spectrum:} [Left] Extra feature obtained in the correlation function due to oscillations in the primordial power spectrum at a characteristic length scale of $r\sim300~Mpc$. [Right] Oscillations in the matter power spectrum on top of the BAO sourced by the primordial oscillations.} 
\end{figure*}

In~\autoref{fig:CorrelationMatterPower} we plot the correlation function (left) and the matter power spectrum (right). Best fit power spectrum from P18TT + HST is used here. Apart from the BAO bump at $r\sim150~Mpc$ we find features around $r\sim300~Mpc$. This is the characteristic wavelength of the oscillations in the {\tt New Spectrum} and the {\tt Restricted Spectrum}. In the matter power spectrum too, we notice oscillations on top of the BAO feature. With ongoing and upcoming large scale structure surveys, such as Dark Energy Spectroscopic Instrument~\cite{DESI}, Euclid~\cite{Euclid} and Vera C. Rubin Observatory~\cite{LSST}, such oscillations can be tested.

\section{Supplementary materials - II: Triangle plots}
Here we provide the triangle plots to demonstrate the correlations between the parameters in detail.
\begin{figure*}[!htb]
\centering
\includegraphics[width=\columnwidth]{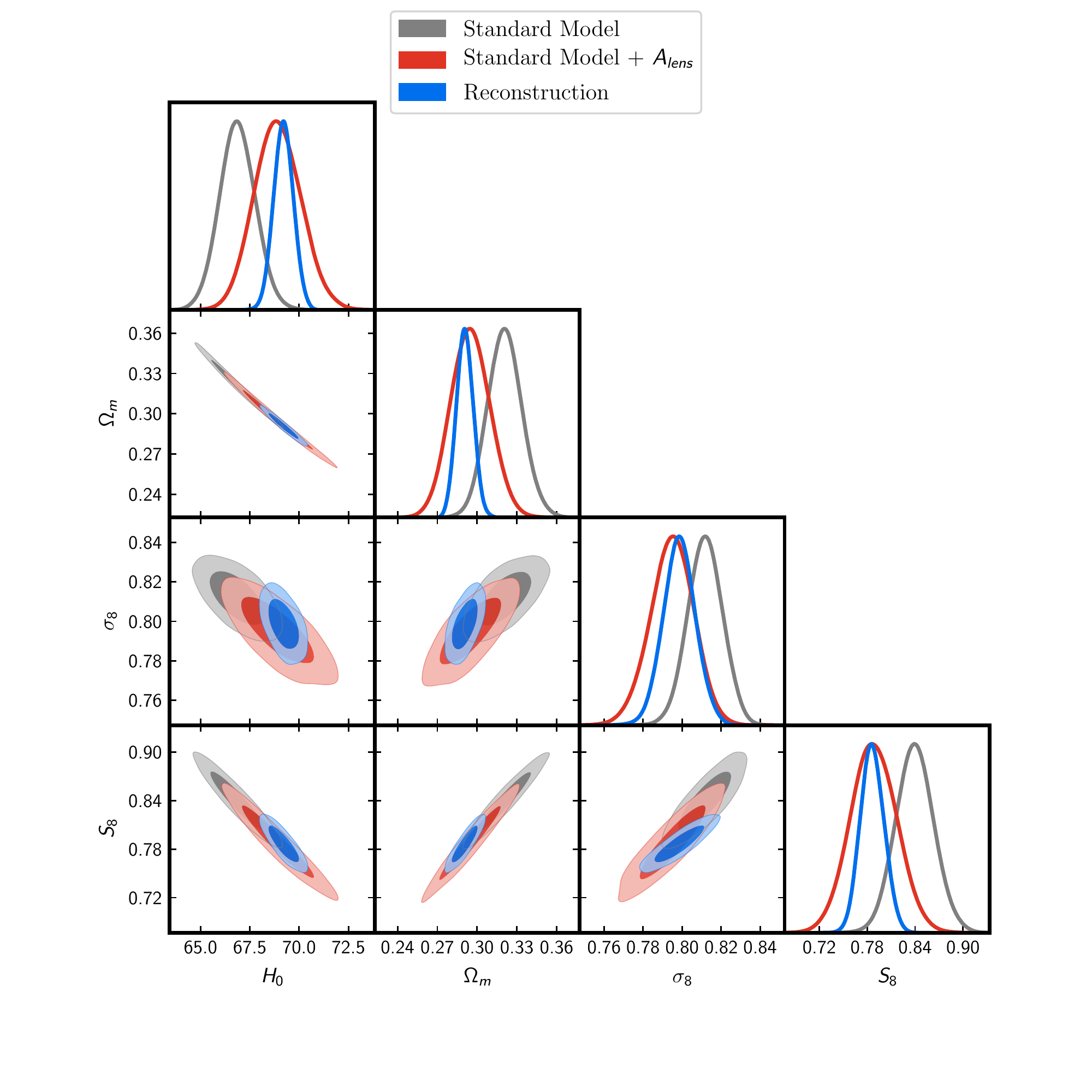}
\caption{\footnotesize\label{fig:ReconAlensTri}Triangle plot corresponding to the {\tt Reconstruction} analysis compared with {\tt Standard Model} and {\tt Standard Model + $A_{lens}$}.}
\end{figure*}
\begin{figure*}[!htb]
\centering
\includegraphics[width=\columnwidth]{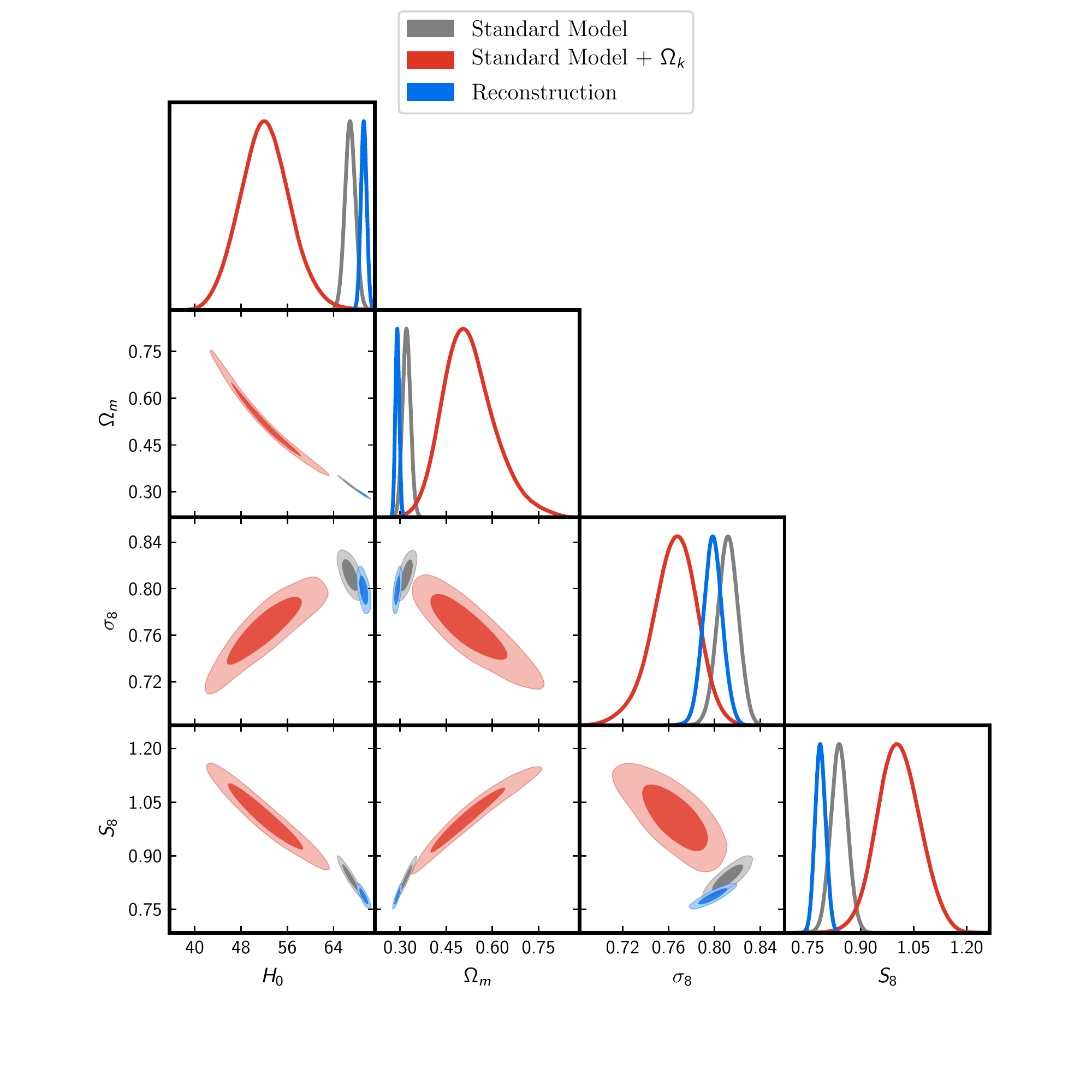}
\caption{\footnotesize\label{fig:ReconOmegaKTri}Triangle plot corresponding to the {\tt Reconstruction} analysis compared with {\tt Standard Model} and {\tt Standard Model + $\Omega_{K}$}.}
\end{figure*}

\begin{figure*}[!htb]
\centering
\includegraphics[width=\columnwidth]{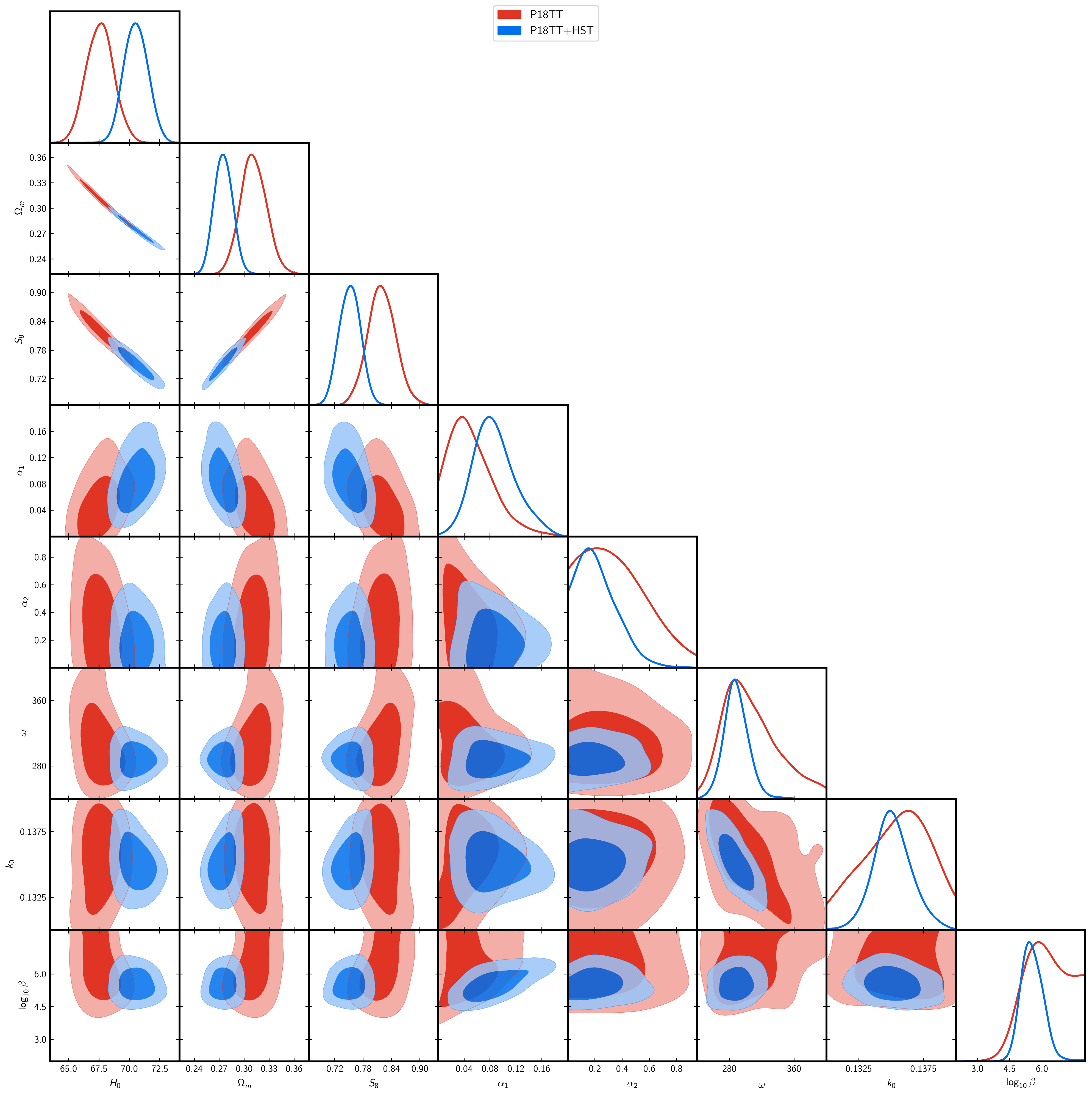}
\caption{\footnotesize\label{fig:NewSpectrumTriangle1}Triangle plot corresponding to the {\tt New Spectrum} analysis.}
\end{figure*}

\begin{figure*}[!htb]
\centering
\includegraphics[width=\columnwidth]{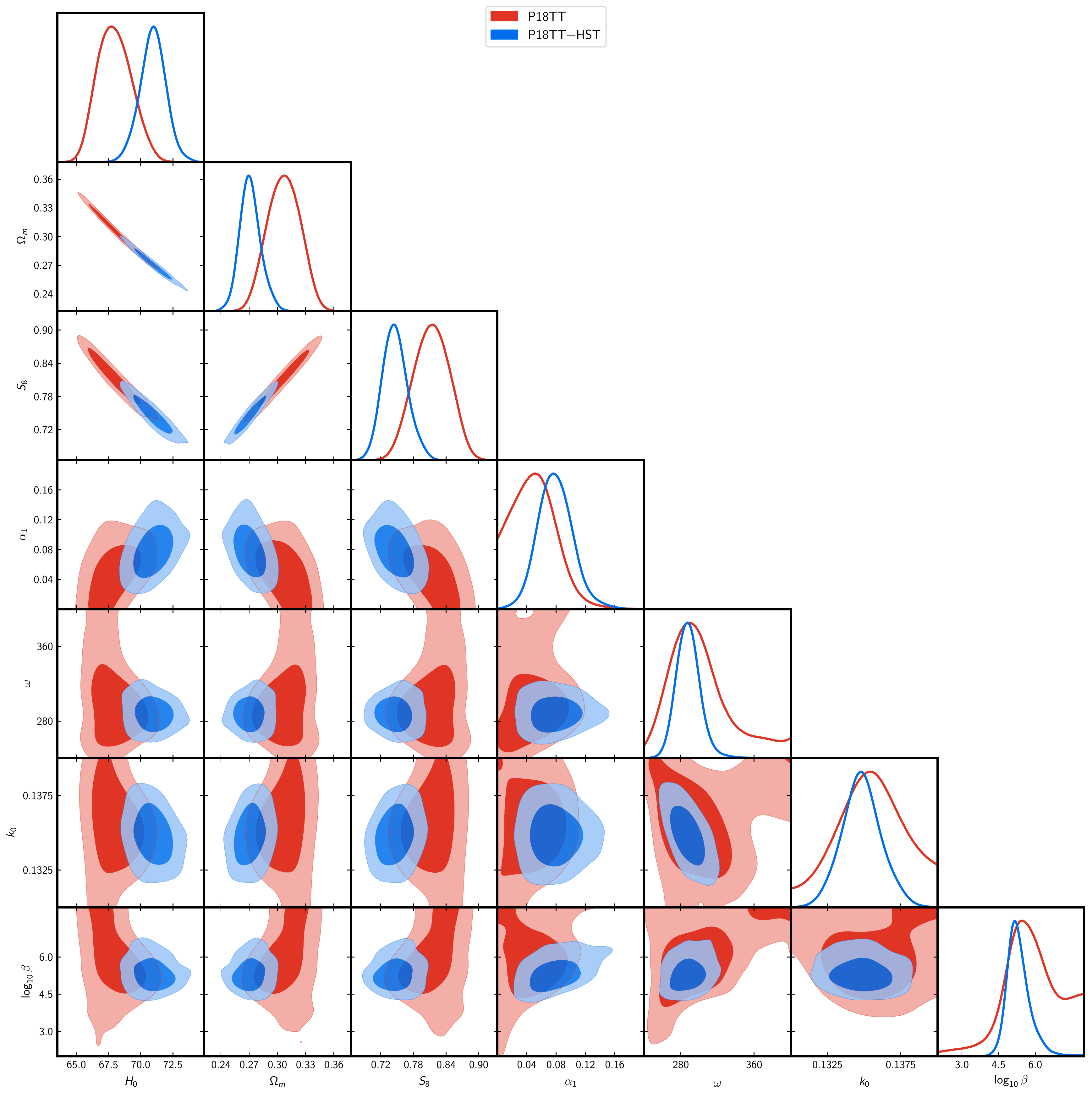}
\caption{\footnotesize\label{fig:NewSpectrumTriangle2}Triangle plot corresponding to the {\tt Restricted Spectrum} analysis.}
\end{figure*}

\begin{figure*}[!htb]
\centering
\includegraphics[width=\columnwidth]{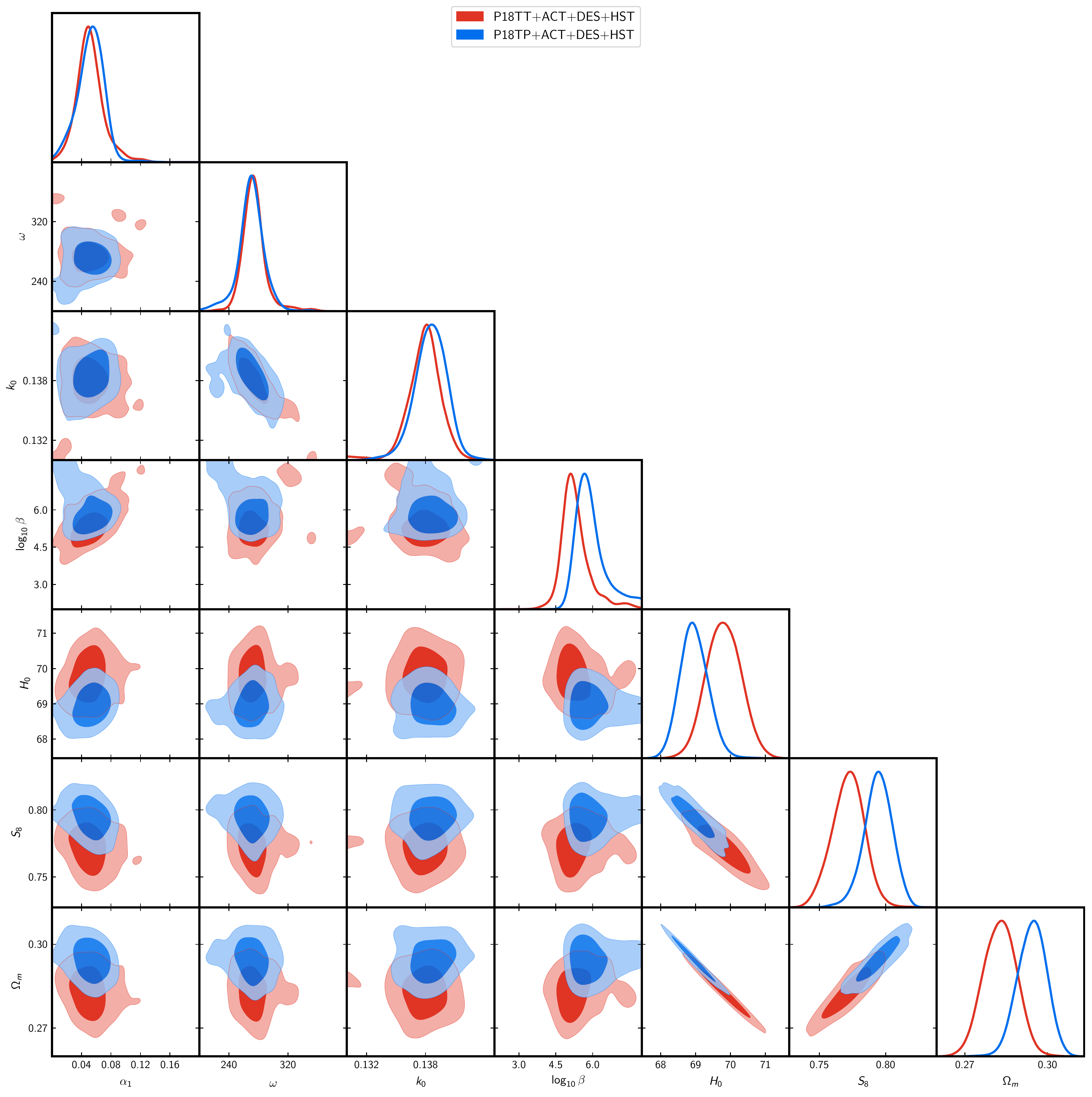}
\caption{\footnotesize\label{fig:ExtendedDataTriangle1}Triangle plot obtained with extended datasets (set 1). Here we have used {\tt Restricted Spectrum} for the analysis.}
\end{figure*}

\begin{figure*}[!htb]
\centering
\includegraphics[width=\columnwidth]{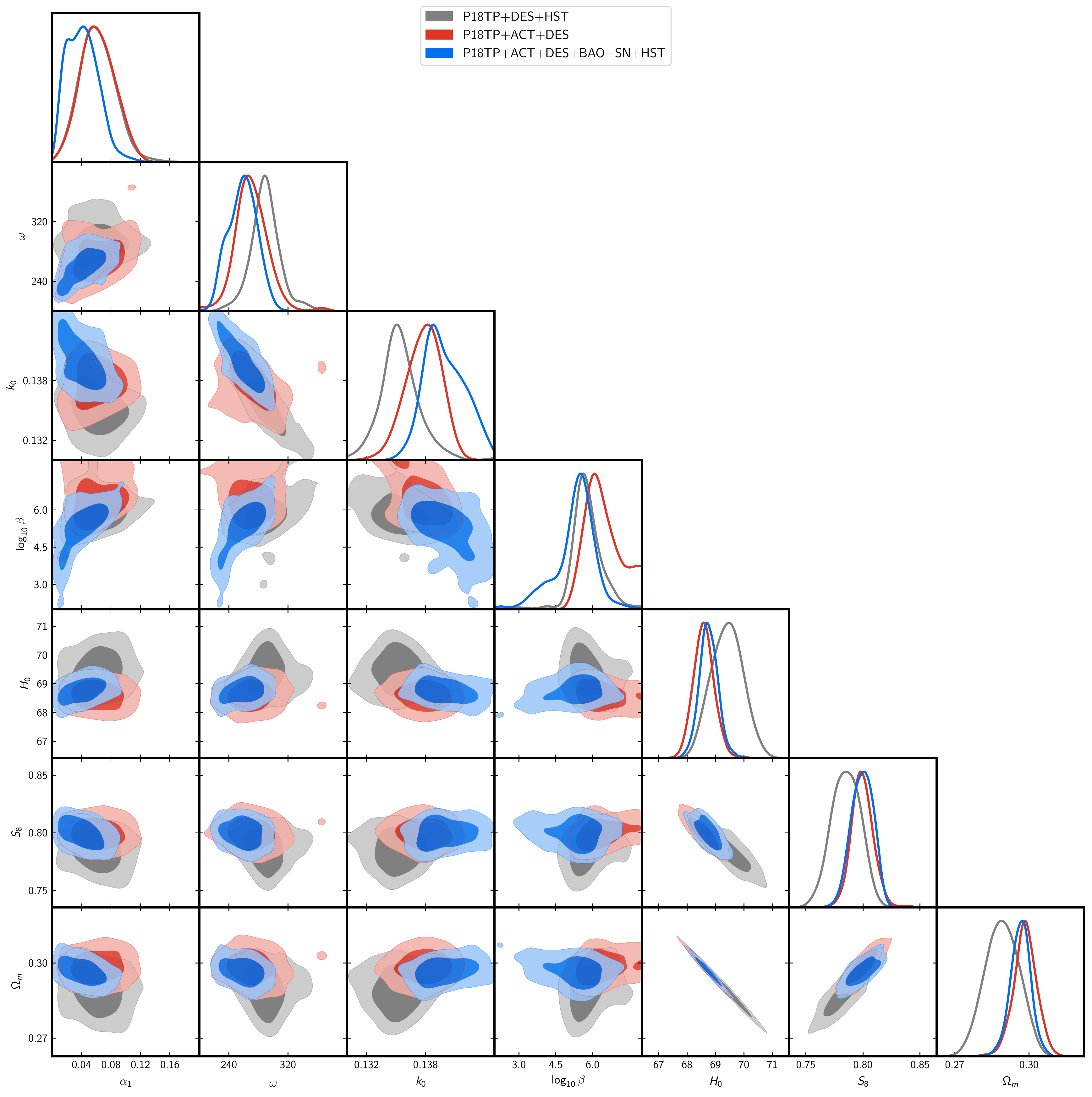}
\caption{\footnotesize\label{fig:ExtendedDataTriangle2}Triangle plot obtained with extended datasets (set 2). Here we have used {\tt Restricted Spectrum} for the analysis.}
\end{figure*}

\end{document}